\documentclass[thmsa,a4paper,11pt]{article}
\usepackage{amsfonts}

\usepackage{amsmath}

\input{tcilatex}

\begin{document}

\setcounter{page}{0} \topmargin0pt \oddsidemargin5mm \renewcommand{%
\thefootnote}{\fnsymbol{footnote}} \newpage \setcounter{page}{0} 
\begin{titlepage}
\begin{flushright}
EMPG-03-10\\
\end{flushright}
\vspace{0.5cm}
\begin{center}
{\Large {\bf Auxiliary matrices for the six-vertex model at $q^N=1$ II.\\
Bethe roots, complete strings and the Drinfeld polynomial} }

\vspace{0.8cm}
{ \large Christian Korff}

\vspace{0.5cm}
{\em School of Mathematics, University of Edinburgh\\
Mayfield Road, Edinburgh EH9 3JZ, U.K.}
\end{center}
\vspace{0.2cm}
 
\renewcommand{\thefootnote}{\arabic{footnote}}
\setcounter{footnote}{0}

\begin{abstract}
The spectra of recently constructed auxiliary matrices for the six-vertex model respectively the 
spin $s=1/2$ Heisenberg chain at roots of unity  are investigated. Two conjectures are formulated
both of which are proven for $N=3$ and are verified numerically for several examples with 
$N>3$. The first conjecture identifies an abelian subset of auxiliary matrices whose eigenvalues 
are polynomials in the spectral variable. The zeroes of these polynomials are shown to fall into 
two sets. One consists of the solutions to the Bethe ansatz equations which determine 
the eigenvalues of the six-vertex transfer matrix. The other set of zeroes contains the 
complete strings which encode the information on the degeneracies of the model 
due to the loop symmetry $\tilde{sl}_2$ present at roots of 1.  The second conjecture then states a polynomial 
identity which relates the complete string centres to the Bethe roots allowing one to determine 
the dimension of the degenerate eigenspaces. Its proof for $N=3$ involves the derivation of 
a new functional equation for the auxiliary matrices and the six-vertex transfer matrix. Moreover, 
it is demonstrated in several explicit examples that the complete strings coincide with the 
classical analogue of the Drinfeld polynomial. The latter is used to classify the 
finite-dimensional irreducible representations of the loop algebra $\tilde{sl}_2$. This suggests 
that the constructed auxiliary matrices not only enable one to solve the six-vertex model but 
also completely characterize the decomposition of its eigenspaces 
into irreducible representations of the underlying loop symmetry.
\medskip
\par\noindent
\end{abstract}
\vfill{ \hspace*{-9mm}
\begin{tabular}{l}
\rule{6 cm}{0.05 mm}\\
C.Korff@ed.ac.uk 
\end{tabular}}
\end{titlepage}
\newpage

\section{Introduction}

This article is a continuation of a previous work \cite{KQ} on auxiliary
matrices for the six-vertex model respectively the $s=1/2$ Heisenberg chain
at roots of 1, 
\begin{equation}
H=\sum_{m=1}^{M}\sigma _{m}^{x}\sigma _{m+1}^{x}+\sigma _{m}^{y}\sigma
_{m+1}^{y}+\frac{q+q^{-1}}{2}\,(\sigma _{m}^{z}\sigma _{m+1}^{z}-1),\quad
\sigma _{M+1}^{x,y,z}\equiv \sigma _{1}^{x,y,z}\;.  \label{H}
\end{equation}
Here $\sigma _{m}^{x,y,z}$ denote the respective Pauli matrices acting on
the $m^{\text{th}}$ site. Throughout this article the deformation parameter $%
q$ will be assumed to be a primitive $N^{\text{th}}$ root of unity with $%
N\geq 3$. While the first part \cite{KQ} of this work was mainly concerned
with the construction of the auxiliary matrices and their geometric
structure, the present paper focuses on the nature of their spectra. This
will allow us to solve the eigenvalue problem of the six-vertex transfer
matrix and thus the Hamiltonian (\ref{H}). In particular, we will derive the
Bethe ansatz equations \cite{Lb67a,Lb67b,Lb67c,St67} 
\begin{equation}
\left\{ \frac{\sinh \frac{1}{2}(u_{j}^{B}+i\gamma )}{\sinh \frac{1}{2}%
(u_{j}^{B}-i\gamma )}\right\} ^{M}=\prod_{\substack{ l=1  \\ l\neq j}}%
^{n_{B}}\frac{\sinh \frac{1}{2}(u_{j}^{B}-u_{l}^{B}+2i\gamma )}{\sinh \frac{1%
}{2}(u_{j}^{B}-u_{l}^{B}-2i\gamma )},\quad q=e^{i\gamma }  \label{BE}
\end{equation}
from representation theory and identify their solutions as zeroes of the
auxiliary matrices' eigenvalues. Furthermore, we will demonstrate in
concrete examples how the spectra of the auxiliary matrices yield
information about the degeneracies connected with the loop symmetry $%
\widetilde{sl}_{2}$ present at roots of 1 \cite{DFM,KM01}. Before we
describe the results of this paper in detail we give a short introduction
into the method of auxiliary matrices and a brief overview of the results
obtained in \cite{KQ}.

\subsection{Auxiliary matrices from quantum group theory}

The concept of auxiliary matrices was originally introduced by Baxter in the
context of his solution to the eight-vertex model \cite
{Bx71,Bx72,Bx73a,Bx73b,Bx73c} and motivated by the lack of
spin-conservation. His approach is described in detail in \cite{BxBook}.
While the six-vertex model preserves the total spin, its
infinite-dimensional symmetry algebra $\widetilde{sl}_{2}$ at roots of unity
does not. It is for this reason that auxiliary matrices provide the
appropriate approach for the discussion of the spectrum and the degenerate
eigenspaces. Unlike the coordinate space Bethe ansatz \cite{B31} they do not
rely on spin-conservation. Also the algebraic Bethe ansatz \cite{QISM} has
serious deficiencies. Away from a root of unity the entries of the monodromy
matrix provide a spectrum generating algebra providing a complete set of
eigenstates. This ceases to be true when $q^{N}=1$ as then certain operator
products in the Yang-Baxter algebra vanish. If one wants to resolve the
structure of the degenerate eigenspaces at roots of unity the concept of
auxiliary matrices is therefore the only method left.

However, as far as the six-vertex model is concerned an explicit
construction for an auxiliary matrix has only been given by Baxter for the
sectors of vanishing total spin, cf. formula (101) in \cite{Bx73a}. His
expression applies to generic values of the deformation parameter. More
recently, auxiliary matrices related to the six-vertex model for $q^{N}\neq
1 $ have been studied using infinite-dimensional representations of quantum
groups \cite{BLZ97,AF97,BLZ99,RW02}. In \cite{RW02} the connection with the
lattice model has been explicitly investigated and Baxter's result has been
extended to spin sectors different from zero involving formal infinite power
series.$\smallskip $

Quantum group theory also played the key role in the construction of
auxiliary matrices at $q^{N}=1$ discussed in \cite{KQ}. It needs to be
emphasized that in contrast to \cite{BLZ97,AF97,BLZ99,RW02} the construction
in \cite{KQ} uses finite-dimensional representations which are special to
the root-of-unity case. Furthermore, this approach differs in several
essential points from the one outlined by Baxter, see Sections 1.2 and 1.3
in \cite{KQ}. Nevertheless, the key idea remains the same.$\smallskip $

Recall that the statistical lattice model is defined in terms of the
transfer matrix $T(z)$ (defined in equation (\ref{T}) below) which besides
the deformation parameter $q$ depends on a spectral variable $z$. The
spin-chain Hamiltonian (\ref{H}) is obtained from the transfer matrix by
taking the logarithmic derivative with respect to $z$ and setting $z=1$
afterwards. One now introduces an additional matrix $Q$ which following
Baxter is called ``auxiliary''. Its defining property is the solution of a
suitable functional relation which allows one to solve the eigenvalue
problem of the transfer matrix $T$ in terms of $Q$. The main result of \cite
{KQ} was to explicitly solve the following operator functional equation, 
\begin{equation}
Q_{p}(z)T(z)=\phi _{1}(z)^{M}Q_{p^{\prime }}(zq^{2})+\phi
_{2}(z)^{M}Q_{p^{\prime \prime }}(zq^{-2})\;.  \label{TQ1}
\end{equation}
Here $\phi _{1},\phi _{2}$ are scalar functions (cf. equation (\ref{phi12})
in this article). The auxiliary matrices $Q_{p},Q_{p^{\prime }},Q_{p^{\prime
\prime }}$ depend on additional complex parameters $p=(\mathbf{x},\mathbf{y},%
\mathbf{z},\mathbf{c}=\mu +\mu ^{-1})\in \mathbb{C}^{4}$ with$\;\mathbf{z}%
\neq 1$, whose appearance is connected with the enhanced symmetry of the
six-vertex model at roots of 1. These parameters define points on the
following three-dimensional complex hypersurface \cite{RA89,CK,CKP,A94} 
\begin{equation}
\limfunc{Spec}Z:\quad \mathbf{x\,y}+q^{N^{\prime }}(\mathbf{z}+\mathbf{z}%
^{-1})=\mu ^{N^{\prime }}+\mu ^{-N^{\prime }},\quad N^{\prime }=\left\{ 
\begin{array}{cc}
N, & \text{if }N\text{ odd} \\ 
N/2, & \text{if }N\text{ even}
\end{array}
\right. \;.  \label{SZ}
\end{equation}
The points $p^{\prime }=(\mathbf{x},q^{N^{\prime }}\mathbf{y},q^{N^{\prime }}%
\mathbf{z},\mu q+\mu ^{-1}q^{-1})$ and $p^{\prime \prime }=(\mathbf{x}%
,q^{N^{\prime }}\mathbf{y},q^{N^{\prime }}\mathbf{z},\mu q^{-1}+\mu ^{-1}q)$
in the functional equation (\ref{TQ1}) are determined by representation
theory. When $N$ is even one has to make the further restriction $\mathbf{x}=%
\mathbf{y}=0$. For details and the explicit definition of the auxiliary
matrices we refer the reader to \cite{KQ}. For a subvariety of $\limfunc{Spec%
}Z$ their definition is given below (see equation (\ref{Qmu})) in order to
keep this article self-contained. All matrices in the functional equation (%
\ref{TQ1}) have been proven to commute with each other \cite{KQ} whence it
can be written in terms of eigenvalues.

There are two main advantages of considering the auxiliary matrix $Q_{p}$ as
the central object instead of the transfer matrix. First, the Bethe roots $%
u_{j}^{B}$ solving (\ref{BE}) can be directly obtained as zeroes of the
auxiliary matrices' eigenvalues 
\begin{equation}
0=Q_{p}(z_{j}^{B})T(z_{j}^{B})=\phi _{1}(z_{j}^{B})^{M}Q_{p^{\prime
}}(z_{j}^{B}q^{2})+\phi _{2}(z_{j}^{B})^{M}Q_{p^{\prime \prime
}}(z_{j}^{B}q^{-2}),\quad z_{j}^{B}=e^{u_{j}^{B}}q^{-1}\;.
\end{equation}
Second and more importantly, $Q_{p}$ breaks the infinite-dimensional
symmetry of the six-vertex model at roots of unity and therefore is in
general non-degenerate. Employing the functional equation (\ref{TQ1}) one
can show that this implies that the eigenvalues of the auxiliary matrices
must contain factors which are $q^{2}$-periodic. Consequently, the
eigenvalues of $Q_{p}$ possess additional zeroes besides the Bethe roots
which are called complete strings \cite{Bx73c,FM01a} 
\begin{equation}
q^{N}=1:\quad Q_{p}(z_{j}^{S}q^{2\ell })=0,\quad \ell =0,1,2,...,N^{\prime
}-1\;.  \label{string}
\end{equation}
The occurrence of these complete strings at finite length $M$ of the
spin-chain is characteristic to the root-of-unity case. Note that in
contrast to the Bethe roots $z_{j}^{B}$ the string centre $z^{S}$ is not
determined via the functional equation (\ref{TQ1}). This freedom is at the
heart of understanding the infinite-dimensional symmetry at roots of unity.
As we will see in this article the structure of the degenerate eigenspaces
of the transfer matrix and the Hamiltonian (\ref{H}) is completely described
by the complete strings.

\subsection{Results and outline of the article}

The investigation of the spectra of the auxiliary matrices $Q_{p}$ for
arbitrary points on the hypersurface (\ref{SZ}) is a quite complicated task
since they form a non-abelian set. That is, for a generic pair $%
p_{1},p_{2}\in \limfunc{Spec}Z$ and any pair of spectral variables $z,w\in 
\mathbb{C}$ the corresponding auxiliary matrices do in general not commute.
Instead one finds \cite{KQ} that in order to ensure $%
[Q_{p_{1}}(z),Q_{p_{2}}(w)]=0$ one has to enforce the relations 
\begin{equation}
\frac{\mathbf{x}_{1}z^{-\varepsilon N^{\prime }}}{1-\mathbf{z}_{1}}=\frac{%
\mathbf{x}_{2}w^{-\varepsilon N^{\prime }}}{1-\mathbf{z}_{2}},\;\frac{%
\mathbf{y}_{1}z^{\varepsilon N^{\prime }}}{1-\mathbf{z}_{1}^{-1}}=\frac{%
\mathbf{y}_{2}w^{\varepsilon N^{\prime }}}{1-\mathbf{z}_{2}^{-1}}\,,\quad
\varepsilon =0,1\;.  \label{kain}
\end{equation}
Note that this is sufficient to guarantee that all operators in (\ref{TQ1})
commute. While the non-abelian character of the auxiliary matrices makes
them more powerful as a symmetry it complicates the calculation of the
eigenvalues, since the eigenvectors may depend on the additional parameters $%
p$, the spectral variable $z$ and the deformation parameter $q$. We will
therefore focus only on an abelian subset for which the eigenvectors
exclusively depend on $q$ and the spectra consist of polynomials in the
spectral variable $z$. This abelian subset is defined in the following
conjecture which we will prove for the case $N=3$ in this article.\medskip

\noindent \textbf{CONJECTURE 1}. \emph{For integer }$N\geq 3$ \emph{consider
the subvariety in the hypersurface (\ref{SZ}) defined by } 
\begin{equation}
p_{\mu }=(0,0,\mu ^{-N^{\prime }},\mu +\mu ^{-1})\in \limfunc{Spec}Z,\quad
\mu \in \mathbb{C}^{\times }\;.  \label{pmu}
\end{equation}
\emph{Denote the corresponding auxiliary matrices by }$Q_{\mu }(z)\equiv
Q_{p_{\mu }}(z)$\emph{. An explicit definition will be given below, cf.
equations (\ref{Qmu}), (\ref{Lmu}) and (\ref{pimu}). Then one has the
commutation relations} 
\begin{equation}
\lbrack Q_{\mu }(z),Q_{\nu }(w)]=0,\quad \mu ,\nu \in \mathbb{C}^{\times
},\quad z,w\in \mathbb{C}\;.  \label{CON1}
\end{equation}
\emph{Because of this relation we refer to the one-parameter subset of
auxiliary matrices }$Q_{\mu }$\emph{\ as ``abelian''.}\medskip

For $N=3$ this conjecture will be shown to be valid by explicitly
constructing the intertwiner of the quantum loop algebra $U_{q}(\widetilde{sl%
}_{2})$ associated with the evaluation representations $\pi _{z}^{p_{\mu
}},\pi _{w}^{p_{\nu }}$ (see definitions (\ref{pimu}), (\ref{pimuev}) in the
text and (\ref{S}) in the appendix). That is, we will demonstrate for $%
q^{3}=1$ that the tensor products $\pi _{z}^{p_{\mu }}\otimes \pi
_{w}^{p_{\nu }}$ and $\pi _{w}^{p_{\nu }}\otimes \pi _{z}^{p_{\mu }}$ are
isomorphic. Numerical calculations have also been performed for $N=4,5,6,8$
and the conjecture has been found to be valid.\smallskip

The conjecture that this assertion holds true for all $N$ is motivated by
the observation that the necessary criteria for the existence of such an
intertwiner -- which are identical with the one shown in (\ref{kain}) -- are
satisfied. These necessary criteria turn out to be sufficient for the
existence when $\mathbf{x}_{i}$, $\mathbf{y}_{i}\neq 0$. The corresponding
intertwiners have first been obtained in \cite{BS90}. When $\mathbf{x}_{i}=%
\mathbf{y}_{i}=0$ the above criteria are obviously trivially satisfied for
any values of $z,w$ and $\mathbf{z}_{1},\mathbf{z}_{2}$. Unfortunately, the
parametrization used in \cite{BS90} does not allow one to take the limit $%
\mathbf{x}_{i},\mathbf{y}_{i}\rightarrow 0$ and to obtain the corresponding
intertwiners for the subvariety (\ref{pmu}). Thus, we need to prove the
existence for this special case which is done by explicit construction in
the appendix of this article for $N=3$. The case of arbitrary $N$ is left to
future work.\smallskip

It follows from their definition given in equation (\ref{Qmu}) below that
each auxiliary matrix\ can be decomposed as 
\begin{equation}
Q_{\mu }(z)=\sum_{m=0}^{M}Q_{\mu }^{(m)}z^{m}
\end{equation}
where the coefficients $Q_{\mu }^{(m)}$ are independent of the spectral
variable $z$. If (\ref{CON1}) holds true all the coefficients commute. In
fact, one has $[Q_{\mu }^{(m)},Q_{\nu }^{(n)}]=0$. Hence, every eigenvalue
of the auxiliary matrix $Q_{\mu }(z)$ can be written in terms of polynomials
with the most general form being 
\begin{equation}
Q_{\mu }(z)=\mathcal{N}_{\mu }\,z^{n_{\infty }}\,P_{B}(z)P_{\mu
}(z)P_{S}(z^{N^{\prime }},\mu )\;.  \label{QP}
\end{equation}
Here $\mathcal{N}_{\mu }=\mathcal{N}_{\mu }(q)$ is a normalization constant
not depending on $z$ and the three polynomials are given by 
\begin{eqnarray}
P_{B}(z) &=&\prod_{j=1}^{n_{B}}(z-z_{j}^{B}),\;  \label{PB} \\
P_{\mu }(z) &=&\prod_{j=1}^{n}(z-z_{j}(\mu )),\;  \label{Pmu} \\
P_{S}(z^{N^{\prime }},\mu ) &=&\prod_{j=1}^{n_{S}}\prod_{\ell \in \mathbb{Z}%
_{N^{\prime }}}(z-z_{j}^{S}(\mu )q^{2\ell
})=\prod_{j=1}^{n_{S}}(z^{N^{\prime }}-z_{j}^{S}(\mu )^{N^{\prime }})\;.
\label{PS}
\end{eqnarray}
The first polynomial $P_{B}$ contains the zeroes $z_{j}^{B}=z_{j}^{B}(q)$ of
the eigenvalue which do not depend on the parameter $\mu $ and for which
there is at least one $\ell \in \mathbb{Z}_{N^{\prime }}$ such that $%
z_{j}^{B}q^{2\ell }$ is not a zero of $P_{B}$. We will show below that these
zeroes are finite solutions of the Bethe ansatz equations (\ref{BE}), whence
the notation. Via the polynomial $P_{B}(z)$ they determine (up to a possible
sign factor) the eigenvalue of the six-vertex transfer matrix associated
with (\ref{QP}), 
\begin{equation}
T(z)=\phi _{1}(z)^{M}q^{n_{\infty }}\frac{P_{B}(zq^{2})}{P_{B}(z)}+\phi
_{2}(z)^{M}q^{-n_{\infty }}\,\frac{P_{B}(zq^{-2})}{P_{B}(z)}\;.  \label{TPB}
\end{equation}
The power $n_{\infty }$ of the monomial in (\ref{QP}) gives the number of
the ``Bethe roots at infinity'', see e.g. \cite{Bx02} and references
therein. This expression refers to the parametrization in (\ref{BE}) with $%
z_{j}^{B}=e^{u_{j}^{B}}q^{-1}\rightarrow 0$. The appearance of ``infinite''
Bethe roots is another feature characteristic to the model at roots of 1. It
signals the breakdown of the familiar formula that the number $n_{B}$ of
Bethe roots is related to the number of down spins in the corresponding
eigenstate.

The second polynomial $P_{\mu }$ accounts for the possibility of zeroes
depending on $q$ and $\mu $. They occur because the additional parameters $p$
shift in the functional equation (\ref{TQ1}). Again we allow only for zeroes 
$z_{j}(\mu )$ for which there is at least one $\ell \in \mathbb{Z}%
_{N^{\prime }}$ such that $z_{j}(\mu )q^{2\ell }$ is not a zero of $P_{\mu }$%
. Using the transformation behaviour of the auxiliary matrices under
spin-reversal we will show that 
\begin{equation}
P_{\mu }(z)=\prod_{j=1}^{n_{B}}(z-z_{j}^{B}\mu ^{2})=\mu ^{2n_{B}}P_{B}(z\mu
^{-2})\;.  \label{PmuPB}
\end{equation}

The last factor $P_{S}$ contains the contribution of the complete $N^{\prime
}$-strings (\ref{string}), where the string centre $z_{j}^{S}$ may or may
not depend on $\mu $. The additional dependence on the deformation parameter 
$q$ is suppressed in the notation. As mentioned before the contribution of
the complete strings encodes the information on the degeneracies connected
with the loop symmetry at roots of unity. In particular, the number $n_{S}$
of complete strings is related to the dimension of the corresponding
degenerate eigenspace of the transfer matrix. In order to arrive at this
result we need to determine the number of possible eigenvalues (\ref{QP}) in
a degenerate eigenspace of the transfer matrix.

In the case $N=3$ this information will be derived from the following
functional equation, 
\begin{equation}
N=3:\quad Q_{\mu }(z)Q_{\nu }(z\nu ^{2}q^{2})=Q_{\mu \nu q}(z\nu ^{2}q^{2}) 
\left[ q^{M}(z-1)^{M}T(zq)+(zq^{2}-1)^{M}\right] \;.  \label{TQ2}
\end{equation}
The above identity relates a product of two complete string contributions (%
\ref{PS}) to a single one, thus imposing severe restrictions on the possible 
$\mu $-dependence of the string centres $z_{j}^{S}(\mu )$. One finds that
only two possibilities are allowed: namely, one has 
\begin{equation}
\text{either\quad }z_{j}^{S}(\mu )=z_{j}^{S}\quad \text{or\quad }%
z_{j}^{S}(\mu )=z_{j}^{S}\mu ^{2}\;.  \label{zS}
\end{equation}
Here $z_{j}^{S}$ denotes the (constant) value of the string centres in the
limit $\mu \rightarrow 1$. These values are fixed in terms of the Bethe
roots via the following remarkable identity which is also deduced from (\ref
{TQ2}), 
\begin{equation}
\lim_{\mu \rightarrow 1}\mathcal{N}_{\mu }P_{S}(z^{N^{\prime }},\mu
)=z^{-n_{\infty }}\sum_{\ell \in \mathbb{Z}_{N^{\prime }}}\frac{q^{2(\ell
+1)n_{\infty }}(zq^{2\ell }-1)^{M}}{P_{B}(zq^{2\ell })P_{B}(zq^{2(\ell +2)})}%
\;.  \label{PD}
\end{equation}
Both results taken together imply that for each eigenvalue (\ref{TPB}) of
the transfer matrix which possesses a fixed number of Bethe roots, finite
and infinite ones, there are $2^{n_{S}}$ possible eigenvalues of the
auxiliary matrix. Since the auxiliary matrices break the
infinite-dimensional symmetry of the six-vertex model as well as
spin-reversal symmetry they are non-degenerate. Thus, the corresponding $%
2^{n_{S}}$ eigenstates yield a basis for the degenerate eigenspace of the
transfer matrix.

Note that the crucial functional equation (\ref{TQ2}) is only valid for $N=3$
and must be modified for $N>3$. Nevertheless, the outcome ought to hold true
in general leading to the formulation of the second conjecture.\medskip

\noindent \textbf{CONJECTURE 2}. \emph{The identity (\ref{PD}) and the
restriction (\ref{zS}) on the }$\mu $\emph{-dependence of the complete
string centres not only holds true for }$N=3$ \emph{but applies to all
primitive roots of unity of order }$N\geq 3$\emph{. This in particular
implies that each eigenvalue of the six-vertex transfer matrix allows for }$%
2^{n_{S}}$\emph{\ eigenvalues of the auxiliary matrix.}\medskip

While we will provide a proof of this identity only for $N=3$ we performed
numerical checks for $N=5,6,8$ verifying the second conjecture also in these
cases. Note also that this assertion coincides with previous results in the
literature \cite{DFM,FM01a,FM01b,BA}\ obtained by numercial calculations and
use of the loop algebra symmetry $\widetilde{sl}_{2}$ which has been
established in the commensurate sectors $2S^{z}=0\func{mod}N$ \cite{DFM,KM01}%
. In this context Fabricius and McCoy suggested in \cite{FM01b} an
expression for the classical analogue of the Drinfeld polynomial \cite{Drin}%
. The latter describes the finite-dimensional irreducible representations of
the loop algebra \cite{CP}.\smallskip

The main result of this article is that the contribution of the complete
strings (\ref{PD}) of the auxiliary matrices constructed in \cite{KQ}
coincides with the proposed expression for the classical analogue of the
Drinfeld polynomial. We will address this point in the last section of this
paper when discussing concrete examples. They suggest that the spectra of
the auxiliary matrices describe the decomposition of the eigenspaces into
irreducible representations of the loop algebra. This is of particular
importance as the auxiliary matrices have been defined for all spin sectors,
while the loop algebra generators have only been constructed for the sectors
where the total spin is a multiple of the order $N$.\smallskip

The outline of the article is as follows. In Section 2 we introduce our
conventions in defining the six-vertex model and review the definition and
properties of the auxiliary matrices. Section 3 discusses the implications
of the functional equation (\ref{TQ1}) for the form of the eigenvalues (\ref
{QP}). Section 4 deals with the transformation under spin-reversal, a
symmetry which is broken by the auxiliary matrices. Section 5 shows that the
eigenvalues of the auxiliary matrices occur always in pairs of opposite
momenta and gives the relation between them. Section 6 contains the proof of
the crucial functional equation (\ref{TQ2}). Section 7 summarizes the
results. In particular, the connection with the representation theory of the
loop algebra $\widetilde{sl}_{2}$ is made.

\section{Definitions}

In order to keep this paper self-contained we briefly recall our conventions
for the definition of the six-vertex model. The transfer matrix is given as
the following trace over an operator product 
\begin{equation}
T(z)=\limfunc{tr}_{0}R_{0M}(z)R_{0M-1}(z)\cdots R_{01}(z)  \label{T}
\end{equation}
with 
\begin{equation}
R=\frac{a+b}{2}\,1\otimes 1+\frac{a-b}{2}\,\sigma ^{z}\otimes \sigma
^{z}+c\,\sigma ^{+}\otimes \sigma ^{-}+c^{\prime }\sigma ^{-}\otimes \sigma
^{+},\quad \sigma ^{\pm }=\frac{\sigma ^{x}\pm i\sigma ^{y}}{2}  \label{R}
\end{equation}
being defined over $\mathbb{C}^{2}\otimes \mathbb{C}^{2}$ in terms of the
Boltzmann weights of the six-allowed vertex configurations, 
\begin{equation}
a=\rho ,\quad b=\rho \,\frac{\left( 1-z\right) q}{1-zq^{2}},\quad c=\rho \,%
\frac{1-q^{2}}{1-zq^{2}},\quad c^{\prime }=c\,z\;.  \label{h}
\end{equation}
For convenience we shall henceforth set the arbitrary normalization factor
to $\rho \equiv 1$. The lower indices in (\ref{T}) indicate on which pair of
spaces the $R$-matrix acts in the $(M+1)$-fold tensor product of $\mathbb{C}%
^{2}$. The explicit dependence on the parameter $q$ is suppressed in the
notation.

The six-vertex transfer matrix possesses a number of finite symmetries given
by the vanishing of the following commutators 
\begin{equation}
\lbrack T(z),S^{z}]=[T(z),\frak{R}]=[T(z),\frak{S}]=0\;,  \label{Sz}
\end{equation}
where the respective operators are defined as follows 
\begin{equation}
S^{z}=\frac{1}{2}\sum_{m=1}^{M}\sigma _{m}^{z},\quad \frak{R}=\sigma
^{x}\otimes \cdots \otimes \sigma ^{x},\text{\quad }\frak{S}=\sigma
^{z}\otimes \cdots \otimes \sigma ^{z}=(-1)^{M/2-|S^{z}|}\;.  \label{SzRS}
\end{equation}
The first operator is the total spin, the second invokes spin-reversal and
the third has eigenvalue $+1$ or $-1$ depending whether the number of down
spins $n$ in a state is even or odd. The finite symmetries and the
properties of the Boltzmann weights (\ref{h}) can be used to derive for
spin-chains of even length, $M=2M^{\prime },$ the following useful relations
of the transfer matrix, 
\begin{eqnarray}
T(z,q^{-1}) &=&T(z^{-1},q),  \notag \\
T(z,-q) &=&\frak{S}T(z,q)=T(z,q)\frak{S},  \notag \\
T(zq^{-2},q) &=&b(z^{-1})^{-M}T(z^{-1},q)^{t}\frak{\;}.  \label{Tlaws}
\end{eqnarray}
Here we have temporarily introduced the explicit dependence on the
deformation parameter $q$ in the notation. The (\ref{Tlaws}) transformation
properties impose restrictions on the spectrum of the transfer matrix.
Henceforth, we shall assume $M$ to be even.

\subsection{A one parameter family of auxiliary matrices}

As explained in the introduction only a subclass of the auxiliary matrices
constructed in \cite{KQ} will be considered, namely the ones associated with
nilpotent representations. In terms of the hypersurface (\ref{SZ}) this
class of auxiliary matrices corresponds to the points (\ref{pmu}). We now
explicitly define this one-parameter family of auxiliary matrices setting 
\begin{equation}
Q_{\mu }(z)=\limfunc{tr}_{_{0}}L_{0M}^{\mu }(z/\mu )L_{0M-1}^{\mu }(z/\mu
)\cdots L_{01}^{\mu }(z/\mu ),\quad L_{0m}^{\mu }\in \limfunc{End}%
(V_{0}\otimes V_{m}),\;\mu \in \mathbb{C}^{\times }\;.  \label{Qmu}
\end{equation}
Here the operators in the trace can be expressed as $2\times 2$ matrices
with operator-valued entries 
\begin{equation}
L^{\mu }=\left( 
\begin{array}{cc}
A_{\mu } & B_{\mu } \\ 
C_{\mu } & D_{\mu }
\end{array}
\right) =A_{\mu }\otimes \sigma ^{+}\sigma ^{-}+B_{\mu }\otimes \sigma
^{+}+C_{\mu }\otimes \sigma ^{-}+D_{\mu }\otimes \sigma ^{-}\sigma ^{+}\;.
\label{Ldec}
\end{equation}
The operators $A_{\mu },B_{\mu },C_{\mu },D_{\mu }\in \limfunc{End}(\mathbb{C%
}^{N^{\prime }})$ are given as 
\begin{eqnarray}
A_{\mu }(w) &=&wq\,\pi ^{\mu }(t)-\pi ^{\mu }(t)^{-1},\quad   \notag \\
B_{\mu }(w) &=&wq(q-q^{-1})\pi ^{\mu }(t)\pi ^{\mu }(f),\;  \notag \\
C_{\mu } &=&\left( q-q^{-1}\right) \pi ^{\mu }(e)\pi ^{\mu }(t)^{-1},\quad  
\notag \\
D_{\mu }(w) &=&wq\,\pi ^{\mu }(t)^{-1}-\pi ^{\mu }(t)  \label{Lmu}
\end{eqnarray}
with the $N^{\prime }\times N^{\prime }$ matrices $\pi ^{\mu }(t),\pi ^{\mu
}(e),\pi ^{\mu }(f)$ defined through the following action on the standard
basis $\{v_{n}\}$ in $\mathbb{C}^{N^{\prime }}$ \cite{RA89,CK} \footnote{%
This parametrization of the auxiliary matrices and the root-of-unity
representation slightly differs from the one used in \cite{RA89,CK}
respectively \cite{KQ}. Instead of using the parameter $\lambda $ (cf
equation (43), Section 2 in \cite{KQ}) it is more favourable to use the
variable $\mu =\lambda ^{-1}q^{-1}$ as this facilitates the identification
when taking the nilpotent limit from a generic cyclic representation $%
p=\varphi (\xi ,\zeta ,\lambda )$.}, 
\begin{eqnarray}
\pi ^{\mu }(t)^{2}v_{n} &=&\mu ^{-1}q^{-2n-1}v_{n},\quad \pi ^{\mu
}(f)v_{n}=v_{n+1}\;,\quad \pi ^{\mu }(f)v_{N^{\prime }-1}=0\;,  \notag \\
\pi ^{\mu }(e)v_{n} &=&\frac{\mu +\mu ^{-1}-\mu q^{2n}-\mu ^{-1}q^{-2n}}{%
(q-q^{-1})^{2}}\;v_{n-1}\;.  \label{pimu}
\end{eqnarray}
The matrices $\pi ^{\mu }(t)^{2},\pi ^{\mu }(e),\pi ^{\mu }(f)$ define a
representation $\pi ^{\mu }$ of the quantum group $U_{q}(sl_{2})$ at $q^{N}=1
$, i.e. they are subject to the relations 
\begin{eqnarray}
\pi ^{\mu }(t)\pi ^{\mu }(e)\pi ^{\mu }(t)^{-1} &=&q\,\pi ^{\mu }(e),  \notag
\\
\pi ^{\mu }(t)\pi ^{\mu }(f)\pi ^{\mu }(t)^{-1} &=&q^{-1}\pi ^{\mu }(t), 
\notag \\
\lbrack \pi ^{\mu }(e),\pi ^{\mu }(f)] &=&\frac{\pi ^{\mu }(t)^{2}-\pi ^{\mu
}(t)^{-2}}{q-q^{-1}}\;.
\end{eqnarray}
The elements generating the centre of the quantum group take the values 
\begin{equation}
\pi ^{\mu }(f)^{N^{\prime }}=\pi ^{\mu }(e)^{N^{\prime }}=0,\quad \pi ^{\mu
}(t)^{2N^{\prime }}=\mu ^{-N^{\prime }},
\end{equation}
and 
\begin{equation}
q\pi ^{\mu }(t)^{2}+q^{-1}\pi ^{\mu }(t)^{-2}+(q-q^{-1})^{2}\pi ^{\mu
}(f)\pi ^{\mu }(e)=\mu +\mu ^{-1}\;.
\end{equation}
There are several advantages of defining the auxiliary matrix in terms of
the representation $\pi ^{\mu }$. As a consequence of the quantum group
relations one has the identity 
\begin{equation}
L_{12}^{\mu }(w/z)L_{13}^{\mu }(w)R_{23}(z)=R_{23}(z)L_{13}^{\mu
}(w)L_{12}^{\mu }(w/z)  \label{YBE}
\end{equation}
which implies that the auxiliary matrix and the transfer matrix commute, 
\begin{equation}
\lbrack Q_{\mu }(w),T(z)]=0\;.  \label{QT0}
\end{equation}
In addition, one derives from the following non-split exact sequence of
evaluation representations $\pi _{w}^{\mu }\;$\cite{KQ} 
\begin{equation}
0\rightarrow \pi _{w^{\prime }}^{\mu q}\hookrightarrow \pi _{w}^{\mu
}\otimes \pi _{z}^{0}\rightarrow \pi _{w^{\prime \prime }}^{\mu
q^{-1}}\rightarrow 0,\quad w=w^{\prime }q^{-1}=w^{\prime \prime }q=z/\mu 
\label{seq1}
\end{equation}
the specialization of the functional equation (\ref{TQ1}) to the subvariety (%
\ref{pmu}) 
\begin{equation}
Q_{\mu }(z)T(z)=\phi _{1}(z)^{M}Q_{\mu q}(zq^{2})+\phi _{2}(z)^{M}Q_{\mu
q^{-1}}(zq^{-2})\;.  \label{TQmu}
\end{equation}
Here $\pi _{z}^{0}$ is the root of unity limit of the two-dimensional
fundamental representation defining the six-vertex model. The corresponding
point on $\limfunc{Spec}Z$ is $p^{0}=(0,0,q^{N^{\prime }},q^{2}+q^{-2})$.
The complex numbers $w,z$ play the role of evaluation parameters
respectively spectral variables. The coefficient functions are given by%
\footnote{%
These coefficients are obtained from the one in \cite{KQ} by setting $\rho
_{\pm }(z)=(zq)^{\frac{1\pm 1}{2}}$ in equation (69), Section 3.} 
\begin{equation}
\phi _{1}(z,q)=b(z,q)q^{-\frac{1}{2}}\quad \text{and\quad }\phi _{2}(z)=q^{%
\frac{1}{2}}\;.  \label{phi12}
\end{equation}
All matrices in the equation (\ref{TQmu}) have been shown to commute, whence
they can be simultaneously diagonalised and the eigenvalues of $T$ can be
expressed in terms of the eigenvalues of the respective auxiliary
matrices.\smallskip 

For $N=3$ we will show that the auxiliary matrices obey the stronger
commutation relations (\ref{CON1}). For $N>3$ we assume that Conjecture 1
holds for the reasons stated in the introduction.\smallskip

Recall from \cite{KQ} that the auxiliary matrices for generic $p\in \limfunc{%
Spec}Z$ break all the finite symmetries of the six-vertex model. However,
the one-parameter family (\ref{Qmu}) which constitutes a subvariety
preserves two of the finite symmetries of the six-vertex transfer matrix,
namely 
\begin{equation}
\lbrack Q_{\mu }(z),S^{z}]=[Q_{\mu }(z),\frak{S}]=0\;.
\end{equation}
Spin-reversal symmetry on the other hand remains broken \cite{KQ}, 
\begin{equation}
\frak{R\,}Q_{\mu }(z)\frak{R}=Q_{\mu ^{-1}}(z\mu ^{-2})=(-zq/\mu )^{M}\frak{%
\,}Q_{\mu }(z^{-1}q^{-2}\mu ^{2})^{t}\;.  \label{QR}
\end{equation}
Employing the $U_{q}(sl_{2})$ algebra automorphism $e\rightarrow
f,\;f\rightarrow e,\;t^{\pm 1}\rightarrow t^{\pm 1},\;q\rightarrow q^{-1}$
one proves the additional relation 
\begin{equation}
Q_{\mu }(z,q)=Q_{\mu }(zq^{2},q^{-1})^{t}  \label{Qq}
\end{equation}
which allows one to derive the adjoint of the auxiliary matrix 
\begin{equation}
Q_{\mu }(z,q)^{\ast }=Q_{\bar{\mu}}(\bar{z},q^{-1})^{t}=Q_{\bar{\mu}}(\bar{z}%
q^{-2},q)\;.  \label{herm}
\end{equation}
These identities will prove crucial in the following investigation of the
eigenvalues (\ref{QP}).

\section{The $TQ$ equation in terms of eigenvalues}

We start our analysis of the spectra of the auxiliary matrices by inserting
the expression (\ref{QP}) in the functional equation (\ref{TQmu}). Recall
from the introduction that (\ref{QP}) is the most general form of the
eigenvalues provided (\ref{CON1}) is true. By abuse of notation we will
denote the operators and their eigenvalues by the same symbols. We obtain
from (\ref{TQmu}), 
\begin{multline*}
T(z)\mathcal{N}_{\mu }\,P_{B}(z)P_{\mu }(z)P_{S}(z^{N^{\prime }},\mu )= \\
\phi _{1}(z)^{M}\mathcal{N}_{\mu q}\,q^{2n_{\infty }}P_{B}(zq^{2})P_{\mu
q}(zq^{2})P_{S}(z^{N^{\prime }},\mu q) \\
+\phi _{2}(z)^{M}\mathcal{N}_{\mu q^{-1}}\,P_{B}(zq^{-2})P_{\mu
q^{-1}}(zq^{-2})P_{S}(z^{N^{\prime }},\mu q^{-1})
\end{multline*}
The eigenvalues of the transfer matrix (\ref{T}) are independent of $\mu $,
whence the polynomials $P_{\mu },P_{\mu q},P_{\mu q^{-1}}$ must cancel on
both sides of the functional equation (\ref{TQmu}) except for a constant
factor $q^{\pm 2n}$. Up to a possible renumeration of the zeroes $z_{j}(\mu )
$, this implies the following relation 
\begin{equation}
z_{j}(\mu q)=z_{j}(\mu )q^{2}\;.
\end{equation}
By the same argument one deduces that the complete string centres $%
z_{j}^{S}(\mu )$ can only differ by powers of $q$ for $\mu \rightarrow \mu
q^{\pm 1}$, 
\begin{equation}
P_{S}(z^{N^{\prime }},\mu )=P_{S}(z^{N^{\prime }},\mu q),\quad \Rightarrow
\quad P_{S}(z^{N^{\prime }},\mu )=P_{S}(z^{N^{\prime }},\mu ^{N})\;.
\label{PS1}
\end{equation}
In addition, we can conclude that the ratios $\mathcal{N}_{\mu q}/\mathcal{N}%
_{\mu },\mathcal{N}_{\mu q^{-1}}/\mathcal{N}_{\mu }$ of the normalization
factors are independent of $\mu $, 
\begin{equation}
\mathcal{N}_{\mu q}/\mathcal{N}_{\mu }=\mathcal{N}_{\mu }/\mathcal{N}_{\mu
q^{-1}}\;.  \label{Nmu1}
\end{equation}
Thus, we deduce the following preliminary form of the eigenvalues of the
transfer matrix 
\begin{equation}
T(z)=\phi _{1}(z)^{M}q^{2n_{\infty }+2n_{B}}\frac{\mathcal{N}_{\mu q}}{%
\mathcal{N}_{\mu }}\frac{P_{B}(zq^{2})}{P_{B}(z)}+\phi
_{2}(z)^{M}q^{-2n_{\infty }-2n_{B}}\frac{\mathcal{N}_{\mu q^{-1}}}{\mathcal{N%
}_{\mu }}\,\frac{P_{B}(zq^{-2})}{P_{B}(z)}\;.  \label{pT}
\end{equation}
This form is ``preliminary'' as we will determine the ratios of the
normalization constants below. Evaluating the left-hand-side of the $TQ$%
-relation at a zero $z=z_{j}^{B}$ we obtain also a preliminary form of the
Bethe ansatz equations, 
\begin{equation}
0=\phi _{1}(z_{j}^{B})^{M}\mathcal{N}_{\mu q}\,q^{2n_{\infty
}}P_{B}(z_{j}^{B}q^{2})+\phi _{2}(z_{j}^{B})^{M}q^{-2n_{\infty }}\mathcal{N}%
_{\mu q^{-1}}\,P_{B}(z_{j}^{B}q^{-2})\;.  \label{pBE}
\end{equation}
These equations ensure that the eigenvalues of the transfer matrix have
residue zero when the limit $z\rightarrow z_{j}^{B}$ is taken. Below we will
see that the zeroes of the polynomial $P_{B}$ in fact coincide with the
finite solutions of the Bethe ansatz equations (\ref{BE}), i.e. they are the
Bethe roots at $q^{N}=1$. Note that it might also happen that the zeroes $%
z_{j}^{B}=1$ and $z_{j}^{B}=q^{-2}$ simultaneously occur. While this
possibility looks problematic in light of the parametrization used in (\ref
{BE}), equation (\ref{pBE}) shows that it does not pose a problem as both
sides of the equation then vanish. Also the limit $z\rightarrow 1$ yielding
the momentum eigenvalue in (\ref{pT}) stays well-defined. We therefore
include these zeroes in the set of Bethe roots.

\section{Transformation under spin-reversal}

We now discuss the implications of spin-reversal. First note that we can
deduce from (\ref{QT0}), (\ref{CON1}) and the integrability of the
six-vertex model, $[T(z),T(w)]=0$, that the auxiliary and transfer matrix
have common eigenvectors which neither depend on the spectral variable $z$
nor on the parameter $\mu $. Let $v=v(q)$ be such a common eigenvector with
eigenvalue (\ref{QP}). Then we find according to (\ref{QR}) that 
\begin{eqnarray}
Q_{\mu }(z)\frak{R}\,v &=&\frak{R\,}Q_{\mu ^{-1}}(z\mu ^{-2})v  \notag
\label{qR} \\
&=&\mathcal{N}_{\mu ^{-1}}\,z^{n_{\infty }}\mu ^{-2n_{\infty }}P_{B}(z\mu
^{-2})P_{\mu ^{-1}}(z\mu ^{-2})P_{S}(z^{N^{\prime }}\mu ^{-2N^{\prime }},\mu
^{-1})\frak{R}\,v\;.
\end{eqnarray}
Furthermore, the above eigenvalue must satisfy the functional relation (\ref
{TQmu}), 
\begin{eqnarray}
Q_{\mu }(z)T(z)\frak{R}\,v &=&[\phi _{1}(z)^{M}Q_{\mu q}(zq^{2})+\phi
_{2}(z)^{M}Q_{\mu q^{-1}}(zq^{-2})]\frak{R}\,v  \notag \\
\frak{R\,}Q_{\mu ^{-1}}(z\mu ^{-2})T(z)\,v &=&\frak{R}[\phi
_{1}(z)^{M}Q_{\mu ^{-1}q^{-1}}(z\mu ^{-2})+\phi _{2}(z)^{M}Q_{\mu
^{-1}q}(z\mu ^{-2})]\,v\;.  \label{TQmuR}
\end{eqnarray}
Here we have exploited that the transfer matrix is invariant under
spin-reversal. In the previous section we verified that the eigenvalues of $%
T(z)$ have $3n_{B}$\thinspace zeroes at $z_{j}^{B},z_{j}^{B}q^{\pm 2}$.
These zeroes correspond to the finite Bethe roots as we will see shortly. In
equation (\ref{TQmuR}) these zeroes must originate from the polynomial 
\begin{equation*}
P_{\mu ^{-1}}(z\mu ^{-2})=\mu ^{-2n}\prod_{j=1}^{n}(z-z_{j}(\mu ^{-1})\mu
^{2})
\end{equation*}
as the factors $P_{B}(z\mu ^{-2})$ and the contribution of the complete
strings $P_{S}(z^{N^{\prime }}\mu ^{-2N^{\prime }},\mu ^{-N})$ cancel on
both sides of the equality sign. In fact, replacing $\mu \rightarrow \mu
^{-1}$ we obtain the expression 
\begin{equation*}
T(z)=\phi _{1}(z)^{M}\frac{\mathcal{N}_{\mu q^{-1}}}{\mathcal{N}_{\mu }}\,%
\frac{P_{\mu q^{-1}}(z\mu ^{2})}{P_{\mu }(z\mu ^{2})}+\phi _{2}(z)^{M}\frac{%
\mathcal{N}_{\mu q}}{\mathcal{N}_{\mu }}\,\frac{P_{\mu q}(z\mu ^{2})}{P_{\mu
}(z\mu ^{2})}
\end{equation*}
We saw already earlier that $T(z)$ does not contain any poles, i.e. the
zeroes $z_{j}(\mu )\mu ^{-2}$ have now to satisfy the preliminary Bethe
ansatz equations, 
\begin{equation*}
0=\phi _{1}(z_{j}(\mu )\mu ^{-2})^{M}\mathcal{N}_{\mu q^{-1}}\,P_{\mu
q^{-1}}(z_{j}(\mu ))+\phi _{2}(z_{j}(\mu )\mu ^{-2})^{M}\mathcal{N}_{\mu
q}\,P_{\mu q}(z_{j}(\mu ))
\end{equation*}
Consequently, we have the zeroes $z_{j}(\mu )\mu ^{-2},z_{j}(\mu )\mu
^{-2}q^{\pm 2}$ of the eigenvalue which must coincide with the zeroes $%
z_{j}^{B},z_{j}^{B}q^{\pm 2}$. Hence, after a possible renumeration we are
lead to the conclusion 
\begin{equation}
z_{j}^{B}=z_{j}(\mu )\mu ^{-2}
\end{equation}
which proves relation (\ref{PmuPB}). Using this identification we can now
determine the ratios of the normalization constants by comparing with the
earlier expression (\ref{pT}) for the transfer matrix eigenvalue and
employing (\ref{Nmu1}), 
\begin{equation}
\left( \frac{\mathcal{N}_{\mu q^{-1}}}{\mathcal{N}_{\mu }}\right)
^{2}=\left( \frac{\mathcal{N}_{\mu }}{\mathcal{N}_{\mu q}}\right)
^{2}=q^{2n_{\infty }+4n_{B}}\;\Rightarrow \;\mathcal{N}_{\mu q}=\pm \mathcal{%
N}_{\mu }q^{-n_{\infty }-2n_{B}}\;.  \label{Nmuq}
\end{equation}
Inserting this result back into (\ref{pT}) the final expression (\ref{TPB})
for the six-vertex transfer matrix eigenvalue associated with (\ref{QP}) is
obtained up to a possible sign factor. This ambiguity is due to the square
root in (\ref{Nmuq}).

The missing step in order to compare this result with the outcome of the
coordinate space Bethe ansatz is to verify that the equations (\ref{pBE})
coincide with (\ref{BE}). Employing (\ref{Nmu1}) a simple calculation gives 
\begin{equation}
\left( \frac{1-z_{j}^{B}q^{2}}{q-z_{j}^{B}q}\right) ^{M}=q^{2n_{\infty
}+2n_{B}-M}\prod_{\substack{ k=1  \\ k\neq j}}^{n_{B}}\frac{%
z_{j}^{B}q^{2}-z_{k}^{B}}{z_{j}^{B}-q^{2}z_{k}^{B}}\;,  \label{BEQ}
\end{equation}
Except for the additional phase in front of the product this coincides with (%
\ref{BE}) if we identify $z=e^{u}q^{-1},\;q=e^{i\gamma }$. For real
eigenvectors we will show momentarily that the phase factor is equal to one.
In the general case of complex eigenvectors we do not have a proof that the
phase factor is always trivial. However, numerical calculations for $%
N=3,4,5,6$ and spin chains up to the length $M=11$ have so far not produced
any counter example.

\section{Pairs of eigenvalues}

In this section we are going to exploit the relations (\ref{QR}) and (\ref
{Qq}) of the auxiliary matrix to show that the eigenvalues occur in pairs.
Combining these two identities we obtain 
\begin{equation}
Q_{\mu }(z,q)=(-zq/\mu )^{M}\frak{\,}Q_{\mu
^{-1}}(z^{-1}q^{-2},q)^{t}=(-zq/\mu )^{M}\frak{\,}Q_{\mu
^{-1}}(z^{-1},q^{-1})\;.  \label{Qt2}
\end{equation}
According to (\ref{CON1}) we can find eigenvectors $v$ which only depend on
the deformation parameter, i.e. $v=v(q)$. Equation (\ref{Qt2}) then shows
that the eigenvectors come either in pairs, $(v(q),v(q^{-1}))$ with 
\begin{equation}
Q_{\mu }(z,q)v(q)=\mathcal{N}_{\mu }(q)\,z^{n_{\infty }}P_{B}(z,q)P_{\mu
}(z,q)P_{S}(z^{N^{\prime }},\mu ^{N})v(q)\;,
\end{equation}
and 
\begin{multline}
Q_{\mu }(z,q)v(q^{-1})=  \notag \\
(-zq/\mu )^{M}\mathcal{N}_{\mu ^{-1}}(q^{-1})\,z^{-n_{\infty
}}P_{B}(z^{-1},q^{-1})P_{\mu ^{-1}}(z^{-1},q^{-1})P_{S}(z^{-N^{\prime }},\mu
^{-N})v(q^{-1}),
\end{multline}
or are real, $v(q)=v(q^{-1})=\overline{v(q)},$ in case of which the two
eigenvalues above must be equal. From (\ref{Tlaws}) we infer that the two
eigenvectors $(v(q),v(q^{-1}))$ have momentum $k$ of opposite sign with 
\begin{equation*}
e^{ik}:=\lim_{z\rightarrow 1}T(z,q)=\pm q^{\frac{M}{2}-n_{\infty
}-2n_{B}}\prod_{j=1}^{n_{B}}\frac{1-z_{j}^{B}(q)q^{2}}{1-z_{j}^{B}(q)}\;.
\end{equation*}
Rewriting the eigenvalue of $v(q^{-1})$ in the form (\ref{QP}) one deduces
that the number of infinite Bethe roots transforms according to 
\begin{equation}
n_{\infty }\rightarrow M-n_{\infty }-2n_{B}-n_{S}N^{\prime },  \label{infB}
\end{equation}
while the finite Bethe roots and complete string centres of the respective
eigenvalues are related by the transformations 
\begin{equation}
z_{j}^{B}(q)\rightarrow 1/z_{j}^{B}(q^{-1})\quad \text{and\quad }%
z_{j}^{S}(\mu ^{N})\rightarrow 1/z_{j}^{S}(\mu ^{N})\;.  \label{zBlaws}
\end{equation}
Finally one finds for the normalization constants the mapping 
\begin{equation}
\mathcal{N}_{\mu }(q)\rightarrow \mathcal{N}_{\mu
^{-1}}(q^{-1})(-1)^{M+n_{S}N^{\prime }}\mu
^{-M-2n_{B}}q^{M}\prod_{j=1}^{n_{B}}z_{j}^{B}(q^{-1})^{2}%
\prod_{j=1}^{n_{S}}z_{j}^{S}(\mu ^{-1})^{N^{\prime }}\;.
\end{equation}

The case of real eigenvectors can only happen when the corresponding
eigenvalues of the translation operator $T(z=1,q)$ are real, i.e. $k=0,\pi $%
. Then (\ref{Qt2}) becomes an identity in terms of eigenvalues and Bethe
roots as well as string centres must be invariant under the transformation
laws (\ref{infB}), (\ref{zBlaws}). This implies for $v(q)=v(q^{-1})$ the
identities 
\begin{gather}
M=2n_{\infty }+2n_{B}+n_{S}N^{\prime },  \label{Msum} \\
P_{B}(z^{-1},q^{-1})=(-z)^{-n_{B}}P_{B}(z,q)%
\prod_{j=1}^{n_{B}}z_{j}^{B}(q)^{-1},  \label{PBinv} \\
P_{S}(z^{-N^{\prime }},\mu ^{-N})=P_{S}(z^{N^{\prime }},\mu
^{N})(-z^{N^{\prime }})^{-n_{S}}\prod_{j=1}^{n_{S}}z_{j}^{S}(\mu
)^{-N^{\prime }}  \label{PSinv}
\end{gather}
Note that (\ref{Msum}) fixes the phase factor in (\ref{BEQ}) to be $%
q^{-n_{S}N^{\prime }}=(-1)^{n_{S}(N+1)}$ and completes the derivation of the
Bethe ansatz equations (\ref{BE}) for odd roots of unity and real
eigenvectors. For $N$ even we will see momentarily that there is always an
even number of complete strings, whence (\ref{BE}) also applies in this case.

The relation (\ref{Msum}) also implies that in a degenerate eigenspace of
the transfer matrix with real eigenvectors and a fixed set of Bethe roots
the number of complete strings is constant. Below we will see for $N=3$ that
this also holds true for complex eigenvectors by proving the functional
equation (\ref{TQ2}) and the identity (\ref{PD}).

Inserting the expressions (\ref{PBinv}) and (\ref{PSinv}) into the identity (%
\ref{Qt2}) yields the following equation for the complete string centres, 
\begin{equation}
(\mathcal{N}_{\mu }/\mathcal{N}_{\mu ^{-1}})\mu
^{M+2n_{B}}\prod_{j=1}^{n_{S}}z_{j}^{S}(\mu )^{N^{\prime
}}=\,q^{2n_{B}}\prod_{j=1}^{n_{B}}(z_{j}^{B})^{2}=1\;.  \label{Nmu2}
\end{equation}
Here we have used that 
\begin{equation}
\lim_{z\rightarrow
1}T(z,q)=(-q)^{-n_{B}}\prod_{j=1}^{n_{B}}z_{j}^{B}(q)^{-1}=\pm 1\;.
\label{zBsum}
\end{equation}
Exploiting that $z_{j}^{S}(\mu ^{-1}q^{\pm 1})^{N^{\prime }}=$ $%
z_{j}^{S}(\mu ^{-1})^{N^{\prime }}$ and (\ref{Nmuq}) leads to the further
restriction 
\begin{equation}
(\mu q)^{M+2n_{B}}(\mathcal{N}_{\mu q}/\mathcal{N}_{\mu
^{-1}q^{-1}})=q^{M-2n_{B}-2n_{\infty }}\mu ^{M+2n_{B}}(\mathcal{N}_{\mu }/%
\mathcal{N}_{\mu ^{-1}})\Rightarrow q^{n_{S}N^{\prime }}=1\;.
\end{equation}
Thus, in the case of even roots of unity there is always an even number of
complete strings. This completes also the derivation of the Bethe ansatz
equations (\ref{BE}) for even roots of unity by showing that the phase
factor in (\ref{BEQ}) is equal to one.

Note that up to this point all relations have been derived for general $%
N^{\prime }\geq 3$. Thus, while the conjecture (\ref{CON1}) is only proven
for $N=3$ in this article, the derivation of the spectrum of the auxiliary
matrices applies to all roots of unity once the commutation relations (\ref
{CON1}) are established.

\section{A new functional equation for $N=3$}

In this section we prove for $q^{3}=1$ the two important formulas (\ref{PD})
and (\ref{zS}) employing the functional equation (\ref{TQ2}). While the
explicit form of this functional equation is characteristic to the case $N=3$
the representation theoretic method applied to derive it is not. Indeed, the
line of argument which employs the decomposition of tensor products of
evaluation representations via exact sequences also applies to the general
case which we leave to future work. We only review the key steps in the
derivation of (\ref{TQ2}), since the strategy is analogous to the one used
in \cite{KQ} to derive (\ref{TQ1}).

We start by recalling the concept of an evaluation representation. The
root-of-unity representation (\ref{pimu}) of the finite quantum algebra $%
U_{q}(sl_{2})$ can be extended to a representation $\pi _{w}^{\mu }$ of the
quantum loop algebra $U_{q}(\widetilde{sl}_{2})$ setting 
\begin{eqnarray}
\pi _{w}^{\mu }(e_{0}) &=&w\,\pi ^{\mu }(f),\quad \pi _{w}^{\mu
}(f_{0})=w^{-1}\pi ^{\mu }(e),\quad \pi _{w}^{\mu }(k_{0})=\pi ^{\mu
}(t)^{-2},  \notag \\
\pi _{w}^{\mu }(e_{1}) &=&\pi ^{\mu }(e),\quad \pi _{w}^{\mu }(f_{1})=\pi
^{\mu }(f),\quad \pi _{w}^{\mu }(k_{1})=\pi ^{\mu }(t)^{2}\;.  \label{pimuev}
\end{eqnarray}
Here $\{e_{i},f_{i},k_{i}\}_{i=0,1}$ denotes the Chevalley-Serre basis of
the quantum loop algebra; see e.g. \cite{KQ} for further details and the
conventions used. Employing the coproduct 
\begin{equation}
\Delta (e_{i})=e_{i}\otimes 1+k_{i}\otimes e_{i},\quad \Delta
(f_{i})=f_{i}\otimes k_{i}^{-1}+1\otimes f_{i},\quad \Delta
(k_{i})=k_{i}\otimes k_{i}  \label{cop}
\end{equation}
one can build tensor products of representations. In the present context we
consider the tensor product $\pi _{w}^{\mu }\otimes \pi _{u}^{\nu }$ of the
evaluation representations associated with (\ref{pimu}). Without loss of
generality we can set $u=1$. The corresponding representation spaces, which
we denote by the same symbol, correspond to the auxiliary spaces of the $Q$%
-matrices on the left hand side of the functional equation (\ref{TQ2}). If
the evaluation parameter $w$ is set to the special value $w=q/\mu \nu $ the
tensor product $\pi _{w}^{\mu }\otimes \pi _{1}^{\nu }$ becomes
decomposable, i.e. it contains subrepresentations $W_{1},W_{2}$ of the
quantum loop algebra giving rise to the non-split exact sequence 
\begin{equation}
0\rightarrow W_{1}\overset{\imath }{\hookrightarrow }\pi _{w}^{\mu }\otimes
\pi _{1}^{\nu }\overset{\tau }{\rightarrow }W_{2}\rightarrow 0,\quad w=q/\mu
\nu \;.  \label{seq2}
\end{equation}
Here $\imath :W_{1}\hookrightarrow \pi _{w}^{\mu }\otimes \pi _{1}^{\nu }$
is the inclusion and $\tau :\pi _{w}^{\mu }\otimes \pi _{1}^{\nu
}\rightarrow W_{2}=\pi _{w}^{\mu }\otimes \pi _{1}^{\nu }/W_{1}$ the
quotient projection. The representations $W_{1},W_{2}$ respectively the maps 
$\imath ,\tau $ need to be determined. This can be achieved by using the
intertwiner $S(w):\pi _{w}^{\mu }\otimes \pi _{1}^{\nu }\rightarrow \pi
_{w}^{\mu }\otimes \pi _{1}^{\nu }$ detailed in the appendix and exploiting
the fact that $\ker S(q/\mu \nu )=W_{1}$. One finds 
\begin{equation}
W_{1}=\pi _{w^{\prime }}^{\mu ^{\prime }}\otimes \pi _{z^{\prime }}^{0}\quad 
\text{and}\quad W_{2}=\pi _{w}^{\mu }\otimes \pi _{1}^{\nu }/W_{1}=\pi
_{w^{\prime \prime }}^{\mu ^{\prime }}  \label{seq2a}
\end{equation}
with the various parameters given by 
\begin{equation}
w=q/\mu \nu ,\quad \mu ^{\prime }=\mu \nu q,\quad w^{\prime }=w^{\prime
\prime }=w\nu q,\quad z^{\prime }=w\mu q\;.  \label{seqdata}
\end{equation}
Here $\pi ^{0}$ is the root of unity limit, $q^{3}\rightarrow 1$, of the
two-dimensional representation of $U_{q}(sl_{2})$ in terms of Pauli matrices
and $\pi _{z^{\prime }}^{0}$ the associated evaluation representation of the
quantum loop algebra. The explicit form of the inclusion and quotient
projection is given in the appendix. The functional equation (\ref{TQ2}) now
follows from the definitions (\ref{T}), (\ref{Qmu}) and the identities 
\begin{eqnarray}
L_{13}^{\mu }(z/\mu )L_{23}^{\nu }(z\nu q^{2})(\imath \otimes 1)
&=&q(z-1)(\imath \otimes 1)L_{13}^{\mu \nu q}(z\mu ^{-1}\nu q)R_{23}(zq), 
\notag \\
(\tau \otimes 1)L_{13}^{\mu }(z/\mu )L_{23}^{\nu }(z\nu q^{2})
&=&(zq^{2}-1)L^{\mu \nu q}(z\mu ^{-1}\nu q)(\tau \otimes 1)
\end{eqnarray}
which can be verified by explicit calculation using the definitions (\ref{R}%
), (\ref{Lmu}) and the results (\ref{inc}), (\ref{tau}) in the appendix.

Expressing the functional equation (\ref{TQ2}) in terms of the eigenvalues (%
\ref{QP}) we infer that the following ratio 
\begin{equation}
\frac{Q_{\mu }(zq^{2})Q_{\nu }(zq\nu ^{2})}{Q_{\mu \nu q}(zq\nu ^{2})}%
=z^{n_{\infty }}P_{B}(zq^{2})P_{B}(zq^{-2})\,\frac{\mathcal{N}_{\mu }%
\mathcal{N}_{\nu }\,P_{S}(z^{N^{\prime }},\mu ^{N})P_{S}(z^{N^{\prime }}\nu
^{2N^{\prime }},\nu ^{N})}{q^{n_{\infty }+2n_{B}}\mathcal{N}_{\mu \nu
q}\,P_{S}(z^{N^{\prime }}\nu ^{2N^{\prime }},\mu ^{N}\nu ^{N})}  \label{QQQ}
\end{equation}
must be independent of the parameters $\mu ,\nu $. Here we have used the
previous results (\ref{PmuPB}) and (\ref{Nmuq}).\ Consequently, we must have
that the ratio 
\begin{equation}
\mathcal{N}_{\mu }\mathcal{N}_{\nu }/\mathcal{N}_{\mu \nu }=\lim_{\mu
\rightarrow 1}\mathcal{N}_{\mu }\equiv \mathcal{N}  \label{N}
\end{equation}
is independent of $\mu ,\nu $. Furthermore, in order that the dependence on $%
\mu ,\nu $ from the complete string contribution cancels one is lead to the
conclusion (\ref{zS}). Namely, the string centre $z_{j}^{S}(\mu )$ does
either not depend on the parameter $\mu $ at all or it just depends on it
via the simple factor $\mu ^{2}$. It follows that the ratio (\ref{QQQ})
simplifies to the expression 
\begin{equation*}
\frac{Q_{\mu }(zq^{2})Q_{\nu }(zq\nu ^{2})}{Q_{\mu \nu q}(zq\nu ^{2})}=%
\mathcal{N}\;z^{n_{\infty }}P_{B}(zq^{2})P_{B}(zq)\,P_{S}(z^{N^{\prime
}},\mu ^{N}=1)\;.
\end{equation*}
Inserting this identity into (\ref{TQ2}) and using the previously derived
formula (\ref{TPB}) for the eigenvalues of the six-vertex transfer matrix
completes the proof of the desired equality (\ref{PD}) for $N=3$.

We conclude this section by noting that the result (\ref{zS}) now also
allows us to derive the $\mu $-dependence of the normalization constant in (%
\ref{QP}) for real eigenvectors. Employing (\ref{Nmu2}) and (\ref{N})
setting $\mu \rightarrow \mu ^{-1/2},\nu \rightarrow \mu $ one arrives at 
\begin{equation}
\mathcal{N}_{\mu }=\mathcal{N}\,\mu ^{-\frac{M}{2}-n_{B}-N^{\prime
}n_{S}^{\prime }},\quad \mu ^{N^{\prime }n_{S}^{\prime
}}=\prod_{j=1}^{n_{S}}[z_{j}^{S}(\mu ^{1/2})/z_{j}^{S}]^{N^{\prime }}\;.
\end{equation}
Here $n_{S}^{\prime }$ is simply the number of exact string centres which
depend on $\mu ^{2}$. The above identity in particular implies 
\begin{equation}
\mathcal{N}_{\mu q}=\mathcal{N}_{\mu }q^{-\frac{M}{2}-n_{B}-N^{\prime
}n_{S}^{\prime }}  \label{Nmuq2}
\end{equation}
which fixes the arbitrary sign factor in (\ref{Nmuq}) and thus in (\ref{TPB}%
).

\section{Discussion}

In this article we have analysed the eigenvalues of the auxiliary matrices (%
\ref{Qmu}) at roots of unity belonging to the abelian subvariety (\ref{pmu}%
). Let us now summarize what we have learnt from their spectra about the
eigenvalues and eigenspaces of the six-vertex model at roots of 1.

Our starting point and motivation \cite{KQ} for employing auxiliary matrices
was the observation that when $q^{N}=1$ the more commonly used approaches
such as the coordinate space \cite{B31} or algebraic Bethe ansatz \cite{QISM}
have serious drawbacks. For example, one derives from the algebraic Bethe
ansatz away from a root of unity the following expression for the
eigenvalues of the transfer matrix (\ref{T}), 
\begin{equation}
q^{N}\neq 1:\quad T(z,q)=b(z,q)^{M}q^{-n_{B}}\prod_{j=1}^{n_{B}}\frac{%
zq^{2}-z_{j}^{B}(q)}{z-z_{j}^{B}(q)}+q^{n_{B}}\prod_{j=1}^{n_{B}}\frac{%
zq^{-2}-z_{j}^{B}(q)}{z-z_{j}^{B}(q)}  \label{ABA}
\end{equation}
with 
\begin{equation}
q^{N}\neq 1:\quad n_{B}=\frac{M}{2}-|S^{z}|\;.  \label{ABA2}
\end{equation}
Here we have set as before $z_{j}^{B}=e^{u_{j}^{B}}q^{-1}$. One often finds
in the literature that this parametrization is used for all real coupling
values even though it breaks down when $q^{N}=1$. This does not mean that
the root of unity limit $q^{N}\rightarrow 1$ in (\ref{ABA}) is ill-defined,
it simply requires the explicit knowledge of the Bethe roots. The latter,
however, are usually not known due to the intricate nature of the Bethe
ansatz equations (\ref{BE}). What can happen in the root of unity limit is
that some of the Bethe roots drop out of the parametrization (\ref{ABA}). As
a concrete example consider the spin-zero sector of the $M=6$ chain when $%
q^{N}\neq 1$. One finds the three Bethe roots 
\begin{equation*}
M=6,\;S^{z}=0:\quad z_{1}^{B}=1,\;z_{2}^{B}=q^{-1},\;z_{3}^{B}=q^{-2}
\end{equation*}
which belong to one of the eigenvalues of the transfer matrix in the
four-dimensional momentum $k=0$ sector. If the limit $q^{3}\rightarrow 1$ is
taken the products over the Bethe roots in (\ref{ABA}) give simply one and
the eigenvalue becomes degenerate with the eigenvalue of the pseudo-vacuum
consisting of the state where all spins point up (down).

Since the Bethe roots are not known in general one needs a parametrization
of the eigenvalues in terms of the finite solutions to the Bethe ansatz
equations (\ref{BE}) when $q^{N}=1$. This parametrization we found to be 
\begin{equation*}
q^{N}=1:\quad T(z)=b(z)^{M}q^{-\frac{M}{2}+n_{\infty }}\prod_{j=1}^{n_{B}}%
\frac{zq^{2}-z_{j}^{B}(q)}{z-z_{j}^{B}(q)}+q^{\frac{M}{2}-n_{\infty
}}\prod_{j=1}^{n_{B}}\frac{zq^{-2}-z_{j}^{B}(q)}{z-z_{j}^{B}(q)}
\end{equation*}
where the number of Bethe roots is not fixed by the total spin in contrast
to (\ref{ABA2}). Instead, we found the sum rule 
\begin{equation}
q^{N}=1:\quad M-2n_{\infty }-2n_{B}=0\func{mod}N  \label{Msum!}
\end{equation}
for real eigenvectors. Numerical examples for the $M=3,4,5,6,8$ chain and $%
N=3,4,5,6$ showed so far that it also extends to complex eigenvectors
provided $n_{B}\neq 0$. The above sum rule also played an important role in
the derivation of the Bethe ansatz equations (\ref{BE}) from the functional
equation (\ref{TQ1}). This derivation involved an additional phase factor
which according to (\ref{Msum!}) is trivial.\smallskip

Note that the difference between $q^{N}\neq 1$ and $q^{N}=1$ is not ``only''
a difference in the number of Bethe roots and a change of the phase factors
in front of the products in (\ref{ABA}). In the root of unity limit also the
eigenstates of the transfer matrix ``re-organize'' into degenerate
eigenspaces across sectors of different spin. The main objective outlined in
the introduction was to investigate the structure of these degenerate
eigenspaces. We will now discuss how this information is encoded in the
complete strings (\ref{PS}). Recall that the complete string contribution in
the eigenvalues of the auxiliary matrices (\ref{Qmu}) is already fixed by
the Bethe root content via the identities (\ref{zS}) and (\ref{PD}). So far
we have only explained how these results determine the dimension of the
degenerate eigenspaces. For several examples we will now explicitly see how
the complete strings characterize the degenerate eigenspaces in terms of
irreducible representations of the loop algebra $\widetilde{sl}_{2}$.

\subsection{The Drinfeld polynomial and complete strings}

Recall that the loop algebra $\widetilde{sl}_{2}$ has been established as a
symmetry of the six-vertex model in the commensurate sectors where the total
spin is a multiple of the order of the root of unity, i.e. $2S^{z}=0\func{mod%
}N$ \cite{DFM,KM01}. In order to make the connection between the spectra of
the auxiliary matrices and the loop algebra we need first to introduce some
facts about its representation theory \cite{CP}.

There are several basis to write down the loop algebra $\widetilde{sl}_{2}$.
The most convenient one for the present purpose is the mode basis obeying
the relations 
\begin{equation}
h_{m+n}=[x_{m}^{+},x_{n}^{-}],\;[h_{m},x_{n}^{\pm }]=\pm 2x_{m+n}^{\pm
},\;[h_{m},h_{n}]=0,\;[x_{m+1}^{\pm },x_{n}^{\pm }]=[x_{m}^{\pm
},x_{n+1}^{\pm }]\;.  \label{sl2}
\end{equation}
The generators $\{x_{m}^{\pm },h_{m}\}_{m\in \mathbb{Z}}$ can be
successively obtained from the Chevalley-Serre basis of the quantum loop
algebra $U_{q}^{\text{res}}(\widetilde{sl}_{2})$ at $q^{N}=1$\ via the
quantum Frobenius homomorphism \cite{CP,KM01} (for simplicity we only
consider $N$ odd) 
\begin{equation}
E_{1}^{(N)}\rightarrow x_{0}^{+},\quad F_{1}^{(N)}\rightarrow
x_{0}^{-},\quad E_{0}^{(N)}\rightarrow x_{1}^{-},\quad
F_{0}^{(N)}\rightarrow x_{-1}^{+},\quad 2S^{z}/N\rightarrow h_{0}
\label{gen}
\end{equation}
and with the action of the quantum group generators given by \cite{DFM,KM01}%
\footnote{%
Here we have used a different convention for the coproduct than in \cite
{DFM,KM01}. However, one analogously proves in this case that the quantum
group generators commute with the six-vertex transfer matrix in the
commensurate sectors.} 
\begin{multline*}
E_{1}^{(N)}(q)=F_{0}^{(N)}(q^{-1})=\frak{R\,}E_{0}^{(N)}(q)\frak{R}=\frak{R\,%
}F_{1}^{(N)}(q^{-1})\frak{R}= \\
\tsum_{1\leq m_{1}<\cdots <m_{N}\leq M}1\otimes \cdots \otimes \underset{%
m_{1}^{\text{th}}}{\sigma ^{+}}\otimes q^{(N-1)\sigma ^{z}}\cdots \otimes 
\underset{m_{2}^{\text{th}}}{\sigma ^{+}}\otimes q^{(N-2)\sigma ^{z}}\cdots
q^{\sigma ^{z}}\otimes \underset{m_{N}^{\text{th}}}{\sigma ^{+}}\otimes
1\cdots \otimes 1\;.
\end{multline*}
Here $\frak{R}$ denotes the spin-reversal operator. As we are only
considering spin-chains of finite length, all representations of the loop
algebra are finite-dimensional and therefore highest weight \cite{CP}. That
is, there exists a highest weight vector $\Omega $ satisfying 
\begin{equation}
x_{n}^{+}\Omega =0,\quad h_{n}\Omega =x_{n}^{+}x_{0}^{-}\Omega
=x_{0}^{+}x_{n}^{-}\Omega =\lambda _{n}\Omega ,\quad \lambda _{n}\in \mathbb{%
C}\;.  \label{Oma}
\end{equation}
All finite-dimensional irreducible highest-weight representations are
isomorphic to tensor products of evaluation representations \cite{CP}. Let $%
\pi ^{s}:sl_{2}\rightarrow \limfunc{End}\mathbb{C}^{2s+1}$ denote the spin $%
s $ representation of the finite subalgebra $sl_{2}=\{x_{0}^{\pm
},h_{0}\}\subset \widetilde{sl}_{2}$. Then define the evaluation
representation $\pi _{a}^{s}:\widetilde{sl}_{2}\rightarrow \limfunc{End}%
\mathbb{C}^{2s+1}$ by setting 
\begin{equation}
\pi _{a}^{s}(x_{0}^{\pm })=\pi ^{s}(x_{0}^{\pm }),\quad \pi _{a}^{s}(x_{\pm
1}^{\mp })=a^{\pm 1}\pi ^{s}(x_{0}^{\mp }),\quad \pi _{a}^{s}(h_{0})=\pi
^{s}(h_{0}),\quad a\in \mathbb{C}\;.
\end{equation}
The information which evaluation representations are contained in the
highest weight representation $\pi _{\Omega }$ is conveniently encoded in
the classical analogue of the Drinfeld polynomial $P_{\Omega }$ according to
the following correspondence \cite{CP}: 
\begin{equation}
\pi _{\Omega }\cong \pi _{a_{1}}^{s_{1}}\otimes \cdots \otimes \pi
_{a_{n}}^{s_{n}}\quad \Leftrightarrow \quad P_{\Omega
}(u)=\prod_{j=1}^{n}(1-a_{j}u)^{2s_{j}}\;.  \label{Poma}
\end{equation}
Here all zeroes $a_{j}$ are different. The Drinfeld polynomial can be
explicitly calculated from the eigenvalues $\lambda _{n}$ of the Cartan
elements $h_{n}$ via the following Laurent series expansions around $u=0$
and $u=\infty $ \cite{CP}, 
\begin{equation}
\sum_{n=0}^{\infty }\lambda _{n}u^{n}=\deg P_{\Omega }-u\frac{P_{\Omega
}^{\prime }(u)}{P_{\Omega }(u)},\quad \quad \sum_{n=1}^{\infty }\lambda
_{-n}u^{-n}=-u\frac{P_{\Omega }^{\prime }(u)}{P_{\Omega }(u)}\;.
\label{Poma0}
\end{equation}

The important observation in connection with the auxiliary matrices
constructed in \cite{KQ} is now the following: the Drinfeld polynomials of
the highest weight representations spanning the degenerate eigenspaces of
the transfer matrix coincide with the complete string contributions (\ref{PD}%
) appearing in the spectrum of $Q_{\mu }(z)$ when we identify $u=z^{N}$.
That is, we find up to a possible renummeration the identification 
\begin{equation}
\dim \pi _{\Omega }=2^{n_{S}}\quad \text{and\quad }a_{j}=\lim_{\mu
\rightarrow 1}z_{j}^{S}(\mu )^{-N}\;.  \label{loopdata}
\end{equation}
At the moment we do not have a general proof of this assertion but we have
verified it for several examples; see the appendix. We consider one of them
in detail to illustrate the interplay between the auxiliary matrices and the
loop algebra $\widetilde{sl}_{2}$.

\subsubsection{Examples for $N=3$}

Consider the spin-chain with $M=6$ sites and the primitive root of unity $%
q=\exp (2\pi i/3)$. Then the vector with all spins up 
\begin{equation}
M=2N=6:\quad \Omega =\uparrow \otimes \uparrow \cdots \otimes \uparrow
\equiv \left| \uparrow \uparrow \cdots \uparrow \right\rangle  \label{oma6}
\end{equation}
lies in the commensurate sector $S^{z}=0\func{mod}N,$ where the loop algebra
generators are defined via (\ref{gen}). The corresponding eigenvalue of the
transfer matrix 
\begin{equation*}
T(z,q)|_{\pi _{\Omega }}=1+b(z,q)^{6}
\end{equation*}
is four-fold degenerate with the eigenspace $\pi _{\Omega }$ spanned by 
\begin{equation*}
\pi _{\Omega }=\{\Omega \}_{S^{z}=3}\oplus \{x_{1}^{-}\Omega
,x_{0}^{-}\Omega \}_{S^{z}=0}\oplus \{\frak{R}\Omega \}_{S^{z}=-3}\;.
\end{equation*}
Here we have indicated via the lower indices the respective spin-sectors.
All eigenvectors have zero momentum. Given the highest weight vector one can
now proceed and calculate the corresponding Drinfeld polynomial. From the
scalar products 
\begin{eqnarray*}
\lambda _{0} &=&\left\langle \Omega |x_{0}^{+}x_{0}^{-}|\Omega \right\rangle
=\left\langle \Omega \left| E_{1}^{(3)}F_{1}^{(3)}\right| \Omega
\right\rangle =\tfrac{1}{2}\dim \pi _{\Omega }=2, \\
\lambda _{1} &=&\left\langle \Omega |x_{0}^{+}x_{1}^{-}|\Omega \right\rangle
=\left\langle \Omega \left| E_{1}^{(3)}E_{0}^{(3)}\right| \Omega
\right\rangle =a_{+}+a_{-}=20, \\
\lambda _{2} &=&\left\langle \Omega |x_{0}^{+}x_{2}^{-}|\Omega \right\rangle
=\left\langle \Omega \left| (x_{0}^{+}x_{1}^{-})^{2}-\tfrac{1}{2}%
(x_{0}^{+})^{2}(x_{1}^{-})^{2}\right| \Omega \right\rangle
=a_{+}^{2}+a_{-}^{2}=398,\;...
\end{eqnarray*}
one finds (see also \cite{D02}) 
\begin{equation}
\,P_{\Omega }(u)=(1-a_{+}u)(1-a_{-}u)\quad \text{with\quad }a_{\pm }=10\pm 3%
\sqrt{11}=(10+3\sqrt{11})^{\pm 1}\quad .  \label{Poma6}
\end{equation}

Diagonalising the auxiliary matrix $Q_{\mu }(z)$ in the respective spin
sectors one computes the following complete string contributions in the
subspace of momentum zero, 
\begin{eqnarray}
S^{z} &=&+3:\quad P_{S}(z^{3},\mu ^{6})=z^{6}-20z^{3}\mu ^{3}+\mu
^{6}=(z^{3}-a_{+}\mu ^{6})(z^{3}-a_{-}\mu ^{6})  \notag \\
S^{z} &=&-3:\quad P_{S}(z^{3},\mu ^{6})=z^{6}\mu ^{6}-20z^{3}\mu
^{3}+1=(z^{3}-a_{+})(z^{3}-a_{-})
\end{eqnarray}
and 
\begin{multline}
S^{z}=0:\quad P_{S}^{\pm }(z^{3},\mu ^{6})=z^{6}-z^{3}\left( 10(\mu
^{6}+1)\pm 3\sqrt{11}(\mu ^{6}-1)\right) +1  \notag \\
=(z^{3}-a_{\pm }\mu ^{6})(z^{3}-a_{\mp })\;.
\end{multline}
Note that $\{E_{0}^{(3)}\Omega ,F_{1}^{(3)}\Omega \}$ are in general not
eigenvectors of the auxiliary matrix, but that the eigenvectors of $Q_{\mu
}(z)$ are contained in the two-dimensional space spanned by them. Taking the
limit $\mu \rightarrow 1$ and identifying $z^{3}=u$ we recover the Drinfeld
polynomial (\ref{Poma6}) from the complete strings. In this limit the
auxiliary matrix becomes degenerate -- as the representation underlying the
definition (\ref{Qmu}) becomes reducible -- and $E_{0}^{(3)}\Omega
,F_{1}^{(3)}\Omega $ are now both eigenvectors of the auxiliary matrix.
However, in general we want to keep the auxiliary matrix non-degenerate and
therefore $\mu $ should be chosen different from one.

The above example also nicely confirms the picture outlined in the
introduction. According to (\ref{zS}) there are $2^{n_{s}}=4$ possible
eigenvalues of the auxiliary matrix in the degenerate eigenspace of the
transfer matrix, all of which we find realized.\smallskip

Note that the match between the complete string centres and the evaluation
parameters is a virtue particular to the auxiliary matrices (\ref{Qmu})
constructed in \cite{KQ}. In the context of the six-vertex model the other
explicit expression in the literature is Baxter's formula (101) in \cite
{Bx73a} (which applies to all coupling values $\gamma \in \mathbb{R}$ but
only to the sectors of vanishing total spin), 
\begin{equation}
S^{z}=0:\quad Q_{\text{Baxter}}(z)_{\alpha _{1}\cdots \alpha _{M}}^{\beta
_{1}\cdots \beta _{M}}=\exp \left( \tfrac{1}{4}i\gamma
\tsum_{m=1}^{M}\tsum_{n=1}^{m-1}\left( \alpha _{m}\beta _{n}-\alpha
_{n}\beta _{m}\right) +\tfrac{1}{4}u\tsum_{m=1}^{M}\alpha _{m}\beta
_{m}\right) .  \label{BQ}
\end{equation}
Here $z=e^{u}q^{-1},\;q=e^{i\gamma }$. Diagonalizing this matrix in the
two-dimensional subspace $\{E_{0}^{(3)}\Omega ,F_{1}^{(3)}\Omega \}$ of the
spin-zero sector we find for each of the two eigenvalues only a single
complete string with string centres $z^{S}=\pm 1$. Thus, for Baxter's
auxiliary matrix neither the degree of the complete string contribution nor
the values of the string centres are in agreement with the data obtained
from the loop algebra.\smallskip

Admittedly, the above example for the $M=6$ chain is quite simple and we
chose it to illustrate the working of the formulas. One might wonder if the
identification of complete strings and the Drinfeld polynomial also applies
when the highest weight vector is a real Bethe eigenstate, i.e. when finite
Bethe roots are present. We have explicitly worked out the following
examples with $q=\exp (2\pi i/3)$: for the $M=5$ chain one finds five
doublets in the $S^{z}=\pm 3/2$ sectors and for the $M=8$ chain there are
eight quartets in the $S^{z}=3,0,-3$ sectors. In all of these cases there is
agreement between the complete strings (\ref{PS}) calculated from the
auxiliary matrices (\ref{Qmu}) and the Drinfeld polynomial (\ref{Poma}). The
results are presented in the appendix. They also show the working of the
identity (\ref{PD}) which yields expressions for the evaluation parameters
of the loop algebra in terms of Bethe roots. Also the Bethe ansatz equations
are recovered by making a Laurent series expansion in (\ref{PD}). This shows
an intimate link between the Bethe ansatz and the representation theory of
the loop algebra.

\subsection{Comparison with the eight-vertex model}

We conclude by mentioning that the identity (\ref{PD}) coincides with the
trigonometric limit of a recent conjecture by Fabricius and McCoy \cite{FM8v}
on the eigenvalues of Baxter's eight-vertex auxiliary matrix constructed in 
\cite{Bx72}. (Note that this auxiliary matrix is different from the one
discussed in \cite{Bx73a,Bx73b,Bx73c} and \cite{BxBook}.) Based on numerical
results for the $M=8$ chain and $N=3,4,6$ they arrive at the elliptic
analogue of the identity (\ref{PD}) by using a functional equation similar
in nature to (\ref{TQ2}), cf. equations (3.12) and (3.10) in \cite{FM8v}.
However, there are some key differences as Baxter's construction procedure
and the one used in \cite{KQ} are not the same. In particular, the
dependence on continuous parameters analogous to (\ref{SZ}) respectively (%
\ref{pmu}) is absent in Baxter's matrix \cite{Bx72}. In the context of the
auxiliary matrices (\ref{Qmu}) we saw that these parameters play a crucial
role in breaking the loop algebra symmetry and the invariance under
spin-reversal. A direct comparison between the auxiliary matrices for the
eight \cite{Bx72,Bx73a,Bx73b,Bx73c} and six-vertex model \cite{KQ} is not
straightforward: taking the trigonometric limit in the eight-vertex
eigenvalues requires the knowledge of the explicit dependence of various
normalization constants and the Bethe roots on the elliptic nome. Further,
investigations are needed to clarify this point.{\small \medskip }

\noindent \textbf{Acknowledgments}. The author would like to thank the
organizers of the workshop ``Recent Advances in Quantum Integrable Systems''
(Annecy, March 2003) where some of the results in this article have been
presented. It is a pleasure to acknowledge discussions with Harry Braden,
Tetsuo Deguchi, Barry McCoy and Robert Weston. This work has been
financially supported by the EPSRC Grant GR/R93773/01.

\appendix

\section{The intertwiner for $\protect\pi _{w}^{\protect\mu }\otimes \protect%
\pi _{1}^{\protect\nu }$ with $N=3$}

In this section we construct the intertwiner for the following tensor
product of evaluation representations $\pi _{w}^{\mu }\otimes \pi _{1}^{\nu
} $ with $N=3$. If this intertwiner exists the auxiliary matrices $Q_{\mu
}(w),Q_{\nu }(w^{\prime })$ must commute, i.e. the conjecture (\ref{CON1})
holds true for $N=3$. The defining equation of the intertwiner $S$ is given
by 
\begin{equation}
S(w)(\pi _{w}^{\mu }\otimes \pi _{1}^{\nu })\Delta (x)=\left[ (\pi _{w}^{\mu
}\otimes \pi _{1}^{\nu })\Delta ^{\text{op}}(x)\right] S(w),\quad x\in U_{q}(%
\widetilde{sl}_{2})\;.  \label{S}
\end{equation}
Here $\pi _{w}^{\mu }$ is the evaluation representation (\ref{pimuev})
obtained from (\ref{pimu}) for $N=3$. The symbols $\Delta ,\Delta ^{\text{op}%
}$ denote the coproduct (\ref{cop}) and the opposite coproduct. The latter
is obtained by permuting the two factors in (\ref{cop}). The defining
equation (\ref{S}) yields a system of algebraic equations for the matrix
elements of the intertwiner. As $S$ commutes with $(\pi _{w}^{\mu }\otimes
\pi _{1}^{\nu })\Delta (k_{i})$ it is convenient to decompose the tensor
product space into the following direct sum 
\begin{equation*}
V=V_{1}\oplus V_{2}\oplus V_{3}
\end{equation*}
where the respective subspaces are spanned by the following basis vectors 
\begin{eqnarray*}
V_{1} &=&\limfunc{span}\{v_{0}\otimes v_{0},v_{1}\otimes v_{2},v_{2}\otimes
v_{1}\}, \\
V_{2} &=&\limfunc{span}\{v_{0}\otimes v_{1},v_{1}\otimes v_{0},v_{2}\otimes
v_{2}\}, \\
V_{3} &=&\limfunc{span}\{v_{0}\otimes v_{2},v_{1}\otimes v_{1},v_{2}\otimes
v_{0}\}\;.
\end{eqnarray*}
Here $v_{i},i=0,1,2$ denotes the standard basis in $\mathbb{C}^{3}$ used in
the definition (\ref{pimu}) of the representation $\pi ^{\mu }$. The
calculation is cumbersome but straightforward and one finds the following
solution up to a common normalization factor, 
\begin{eqnarray*}
S|_{V_{1}} &=&\left( 
\begin{array}{ccc}
1 & 0 & 0 \\ 
0 & \frac{q(w\mu -\nu )(w\mu \nu -q)}{(wq-\mu \nu )(wq^{2}-\mu \nu )} & 
\frac{(w\mu \nu -q)(\mu ^{2}-q^{2})}{(wq-\mu \nu )(wq^{2}-\mu \nu )} \\ 
0 & \frac{w(w\mu \nu -q^{2})(\nu ^{2}-q^{2})}{(wq-\mu \nu )(wq^{2}-\mu \nu )}
& \frac{q(w\nu -\mu )(w\mu \nu -q)}{(wq-\mu \nu )(wq^{2}-\mu \nu )}
\end{array}
\right) , \\
S|_{V_{2}} &=&\left( 
\begin{array}{ccc}
\frac{q^{2}(w\mu -\nu )}{wq-\mu \nu } & \frac{q-\mu ^{2}}{wq-\mu \nu } & 0
\\ 
\frac{w(q-\nu ^{2})}{wq-\mu \nu } & \frac{q^{2}(w\nu -\mu )}{wq-\mu \nu } & 0
\\ 
0 & 0 & \frac{(w\mu \nu -q)(w\mu \nu -q^{2})}{(wq-\mu \nu )(wq^{2}-\mu \nu )}
\end{array}
\right) , \\
S|_{V_{3}} &=&\left( 
\begin{array}{ccc}
\frac{q^{2}(w\mu -\nu )(w\mu -\nu q^{2})}{(wq-\mu \nu )(wq^{2}-\mu \nu )} & 
\frac{(q-\mu ^{2})(w\mu -\nu )}{(wq-\mu \nu )(wq^{2}-\mu \nu )} & \frac{(\mu
^{2}-q)(\mu ^{2}-q^{2})}{(wq-\mu \nu )(wq^{2}-\mu \nu )} \\ 
\frac{w(w\mu -\nu )(\nu ^{2}-q^{2})}{(wq-\mu \nu )(wq^{2}-\mu \nu )} & \frac{%
\mu \nu (1+q^{2}w^{2})+wq(\mu ^{2}+1)(\nu ^{2}+1)}{(wq-\mu \nu )(wq^{2}-\mu
\nu )} & \frac{(\mu ^{2}-q^{2})(w\nu -\mu )}{(wq-\mu \nu )(wq^{2}-\mu \nu )}
\\ 
\frac{w^{2}(\nu ^{2}-q)(\nu ^{2}-q^{2})}{(wq-\mu \nu )(wq^{2}-\mu \nu )} & 
\frac{w(q-\nu ^{2})(w\nu -\mu )}{(wq-\mu \nu )(wq^{2}-\mu \nu )} & \frac{%
q^{2}(w\nu -\mu )(w\nu -\mu q^{2})}{(wq-\mu \nu )(wq^{2}-\mu \nu )}
\end{array}
\right) \;.
\end{eqnarray*}
We can now use this solution in order to explore the decomposition of the
tensor product at special values of the evaluation parameter $w$. One can
explicitly verify that for $w=q/\mu \nu $ the intertwiner has a non-trivial
kernel consisting of the following six-dimensional space 
\begin{eqnarray*}
\ker S_{1}(q/\mu \nu ) &=&\limfunc{span}\{v_{1}\otimes v_{2},v_{2}\otimes
v_{1}\}, \\
\ker S_{2}(q/\mu \nu ) &=&\limfunc{span}\{v_{2}\otimes v_{2},\tfrac{(\mu
^{2}-q)\nu }{1-q^{2}\nu ^{2}}v_{0}\otimes v_{1}+v_{1}\otimes v_{0}\}, \\
\ker S_{3}(q/\mu \nu ) &=&\limfunc{span}\{\tfrac{\nu ^{2}(\mu ^{2}-q)(\mu
^{2}-q^{2})}{(q-\nu ^{2})(\nu ^{2}q-1)}v_{0}\otimes v_{2}+v_{2}\otimes v_{0},%
\tfrac{\nu (q-\mu ^{2})}{\nu ^{2}q-1}v_{0}\otimes v_{2}+v_{1}\otimes
v_{1}\}\;.
\end{eqnarray*}
This kernel can be identified as a submodule $W_{1}\subset \pi _{w}^{\mu
}\otimes \pi _{1}^{\nu }$ of the quantum loop algebra.

\subsection{The inclusion $\imath :\protect\pi _{w^{\prime }}^{\protect\mu
^{\prime }}\otimes \protect\pi _{z^{\prime }}^{0}\hookrightarrow
W_{1}\subset \protect\pi _{w}^{\protect\mu }\otimes \protect\pi _{1}^{%
\protect\nu }$}

We define the module $W_{1}$ simply by stating the inclusion of the basis
vectors spanning the tensor product $\pi _{w^{\prime }}^{\mu ^{\prime
}}\otimes \pi _{z^{\prime }}^{0}$ into the tensor product $\pi _{w}^{\mu
}\otimes \pi _{1}^{\nu }$. Denote by $\{v_{i}^{\prime }\}$ the basis vectors
in $\pi _{w^{\prime }}^{\mu ^{\prime }}$ and by $\{\uparrow ,\downarrow \}$
the basis vector of the two-dimensional representation of $U_{q}(sl_{2})$.
Then the inclusion $\imath $ is defined by linear extension from the
following relations involving the basis vectors, 
\begin{eqnarray}
\imath \left( v_{2}^{\prime }\otimes \downarrow \right) &=&\alpha
\,v_{2}\otimes v_{2},\quad \imath \left( v_{2}^{\prime }\otimes \uparrow
\right) =\alpha \,v_{2}\otimes v_{1},\quad  \label{inc} \\
\imath \left( v_{1}^{\prime }\otimes \uparrow \right) &=&\gamma
_{0}\,\left\{ \gamma _{1}\,v_{1}\otimes v_{1}+v_{2}\otimes v_{0}\right\}
,\quad \imath \left( v_{1}^{\prime }\otimes \downarrow \right) =\gamma
_{2}\,v_{1}\otimes v_{2}+\gamma _{3}\,v_{2}\otimes v_{1},  \notag \\
\imath \left( v_{0}^{\prime }\otimes \uparrow \right) &=&\beta _{0}\text{%
\thinspace }v_{0}\otimes v_{1}+v_{1}\otimes v_{0},\quad \imath \left(
v_{0}^{\prime }\otimes \downarrow \right) =\beta _{0}\,v_{0}\otimes
v_{2}+\beta _{1}\,v_{1}\otimes v_{1}+\beta _{2}\,v_{2}\otimes v_{0}\;. 
\notag
\end{eqnarray}
The coefficients in the above linear combinations are given by 
\begin{eqnarray*}
\alpha &=&\frac{(\mu ^{2}\nu ^{2}-1)(1-q\mu ^{2}\nu ^{2})}{\nu (\mu
^{2}q-1)(\nu ^{2}-q)}, \\
\beta _{0} &=&\frac{(\mu ^{2}-q)\nu }{1-q^{2}\nu ^{2}},\quad \beta _{1}=\nu
\beta _{0}+1-\gamma _{0}\gamma _{1}q^{2},\quad \beta _{2}=\nu q-\gamma
_{0}q^{2}, \\
\gamma _{0} &=&\frac{\alpha (1-\nu ^{2}q^{2})}{\mu ^{2}\nu ^{2}-1},\quad
\gamma _{1}=\frac{\nu (\mu ^{2}-q^{2})}{q-\nu ^{2}},\quad \gamma _{2}=(\beta
_{0}q\nu -1),\quad \\
\gamma _{3} &=&[q\alpha +\nu (q^{2}-\nu \beta _{0})\;.
\end{eqnarray*}
Acting with the quantum group generators according to (\ref{cop}) on the
basis vectors in the respective tensor products of evaluation
representations one verifies that the above inclusion is well-defined.

\subsection{The projection $\protect\tau :\protect\pi _{w}^{\protect\mu
}\otimes \protect\pi _{1}^{\protect\nu }\rightarrow W_{2}=\protect\pi _{w}^{%
\protect\mu }\otimes \protect\pi _{1}^{\protect\nu }/W_{1}$}

Having identified the submodule $W_{1}$ it remains to verify that the
quotient space $W_{2}$ defines the evaluation representation $\pi
_{w^{\prime \prime }}^{\mu ^{\prime \prime }}$ as outlined in (\ref{seq2}), (%
\ref{seq2a}) and (\ref{seqdata}). This follows when identifying the
equivalence classes of the following vectors in $\pi _{w}^{\mu }\otimes \pi
_{1}^{\nu }$ with the basis vectors $\{v_{i}^{\prime \prime }\}$ in $\pi
_{w^{\prime \prime }}^{\mu ^{\prime \prime }},$%
\begin{eqnarray}
v_{0}^{\prime \prime } &\equiv &\tau (v_{0}\otimes v_{0}),\quad  \notag \\
v_{1}^{^{\prime \prime }} &\equiv &\tau (v_{0}\otimes v_{1}+\nu
q\,v_{1}\otimes v_{0}),\;  \notag \\
v_{2}^{\prime \prime } &\equiv &\tau (v_{0}\otimes v_{2}-\nu
q^{2}\,v_{1}\otimes v_{1}+\nu ^{2}q^{2}\,v_{2}\otimes v_{0})\;.  \label{tau}
\end{eqnarray}
This concludes the proof of the decomposition (\ref{seq2}). Using the
explicit form of the inclusion and projection map one is now in the position
to proof the functional equation (\ref{TQ2}) as described in the text.

\section{Calculation of the Drinfeld polynomial}

We present several examples of calculating the evaluation parameters (\ref
{Poma}) of the loop algebra in the degenerate eigenspaces of the transfer
matrix when $q=\exp (2\pi i/3)$. We then compare the outcome with the
expression (\ref{PD}) derived from the complete strings of the auxiliary
matrices.

\subsection{$M=5,\;S^{z}=\pm 3/2$}

There are in total five doublets for the $M=5$ chain in the sectors $%
S^{z}=\pm 3/2$. The corresponding highest weight vectors $\Omega _{k}$ can
be labelled by their momenta and are defined as follows, 
\begin{equation*}
\Omega _{k}=\sum_{n=1}^{5}e^{ink}T(1,q)^{n}\left| \uparrow \uparrow \uparrow
\uparrow \downarrow \right\rangle ,\quad k/\pi =0,\pm 2/5,\pm 4/5\;.
\end{equation*}
Since there are only doublets occurring in this example the corresponding
Drinfeld polynomials $P_{\Omega _{k}}$ defined in (\ref{Poma}) contain only
one factor. For each highest weight vector the corresponding evaluation
parameter $a(k)$ is calculated using the action of the loop algebra, 
\begin{eqnarray}
x_{0}^{+}x_{1}^{-}\Omega _{k}
&=&[4+3q\,e^{-ik}+(1+2q^{2})\,e^{-2ik}+(1+2q)\,e^{-3ik}+3q^{2}\,e^{-4ik}]\,%
\Omega _{k}  \notag \\
&=&a(k)\Omega _{k}\;.
\end{eqnarray}
In order to compare this result with the complete strings we may either
directly diagonalise the auxiliary matrices (\ref{Qmu}) in the respective
spin-sectors or use the identity (\ref{PD}). In the latter approach one
first solves the Bethe ansatz equation 
\begin{equation}
1=q^{5}\left( \frac{1-z_{B}q^{2}}{1-z_{B}}\right) ^{5}  \label{BE5}
\end{equation}
and then computes from the Bethe roots the complete strings in the limit $%
\mu \rightarrow 1$, 
\begin{equation}
\mathcal{N}(z^{3}-z_{S}^{3})=\sum_{\ell \in \mathbb{Z}_{3}}\frac{(zq^{2\ell
}-1)^{5}}{(zq^{2\ell }-z_{B})(zq^{2(\ell +2)}-z_{B})}=-\frac{3}{z_{B}^{2}}%
\left( 1+\frac{1-10z_{B}^{2}(z_{B}+q^{2})}{z_{B}^{3}}\,z^{3}\right) \;.
\label{PD5}
\end{equation}
Bethe roots and momenta can be easily matched by taking the limit $%
z\rightarrow 1$ in (\ref{TPB}) yielding the following second identity for
the evaluation parameter 
\begin{equation}
a(k)=10+10q^{2}/z_{B}-1/z_{B}^{3},\quad e^{ik}=-q^{\frac{1}{2}}\frac{%
1-z_{B}q^{2}}{1-z_{B}}\;.
\end{equation}
Note that the Bethe ansatz equations (\ref{BE5}) are recovered from the
Laurent series expansion in (\ref{PD5}) by setting all coefficients of the
terms with powers greater than three equal to zero.

\subsection{$M=8,\;S^{z}=3,0,-3$}

For the $M=8$ chain one proceeds similar as in the previous case. One now
has eight quartets whose highest weight vectors in the $S^{z}=3$ sector are
again labelled by their momenta 
\begin{equation*}
\Omega _{k}=\sum_{n=1}^{8}e^{ink}T(1,q)^{n}\left| \downarrow \uparrow
\uparrow \uparrow \uparrow \uparrow \uparrow \uparrow \right\rangle \;.
\end{equation*}
The degree of Drinfeld polynomial $P_{\Omega _{k}}$ is now two, i.e. there
are two evaluation parameters $a_{\pm }=a_{\pm }(k)$ to compute. After some
cumbersome computations one obtains 
\begin{multline*}
a_{+}+a_{-}=\left\langle \Omega _{k}|x_{0}^{+}x_{1}^{-}|\Omega
_{k}\right\rangle =35+15qe^{-ik}+5i\sqrt{3}q^{2}e^{-2ik}-(5-2q)e^{-3ik} \\
+6e^{-4ik}-(5-2q^{2})e^{-5ik}-5i\sqrt{3}qe^{-6ik}+15q^{2}e^{-7ik}
\end{multline*}
and 
\begin{eqnarray*}
4a_{+}a_{-} &=&\left\langle \Omega
_{k}|(x_{0}^{+})^{2}(x_{1}^{-})^{2}|\Omega _{k}\right\rangle \\
&=&4(e^{-ik}+qe^{-2ik}+q^{2}e^{-3ik}+e^{-4ik}+qe^{-5ik}+q^{2}e^{-6ik}+e^{-7ik})^{2}\;.
\end{eqnarray*}
Again we can compare this result against the complete string by
diagonalising the auxiliary matrices or by employing the identity (\ref{PD}%
). In either case one finds upon matching string centres and evaluation
parameters the following expression in terms of the Bethe roots 
\begin{equation*}
a_{+}+a_{-}=56+28q^{2}/z_{B}-1/z_{B}^{3},\quad
a_{+}a_{-}=28+56q^{2}/z_{B}-56/z_{B}^{3}-28q^{2}/z_{B}^{4}+1/z_{B}^{6}\;.
\end{equation*}
Here Bethe roots and momenta are related by the identity 
\begin{equation*}
e^{ik}=q^{2}\frac{1-z_{B}q^{2}}{1-z_{B}}\;.
\end{equation*}
In order to facilitate the comparison we have summarized the results in the
table below.\medskip

\begin{center}
\begin{tabular}{|l|l|l|}
\hline\hline
momentum & string centres/evaluation parameters$\;a_{\pm }$ & Bethe root $%
z^{B}/q^{2}$ \\ \hline\hline
$k=0$ & $\dfrac{1}{2}\left( 29\pm 3\sqrt{93}\right) $ & $-1$ \\ \hline\hline
$k=\pi $ & $\dfrac{1}{2}\left( 83\pm 9\sqrt{85}\right) $ & $1$ \\ 
\hline\hline
$k=\pi /2$ & $\dfrac{1}{2}\left( 13(2+\sqrt{3})\pm \sqrt{165(7+4\sqrt{3})}%
\right) $ & $-2-\sqrt{3}$ \\ \hline\hline
$k=-\pi /2$ & $\dfrac{1}{2}\left( 13(2-\sqrt{3})\pm \sqrt{165(7-4\sqrt{3})}%
\right) $ & $-2+\sqrt{3}$ \\ \hline\hline
$k=\pi /4$ & $a_{+}=38.971...,\quad a_{-}=0.0680614...$ & $1-\frac{3}{\sqrt{2%
}}-\sqrt{\frac{3}{2}(2-2\sqrt{2})}$ \\ \hline\hline
$k=3\pi /4$ & $a_{+}=59.9864...,\quad a_{-}=0.615865...$ & $1+\frac{3}{\sqrt{%
2}}+\sqrt{\frac{3}{2}(2+2\sqrt{2})}$ \\ \hline\hline
$k=5\pi /4$ & $a_{+}=1.62373...,\quad a_{-}=0.0166705...$ & $1+\frac{3}{%
\sqrt{2}}-\sqrt{\frac{3}{2}(2+2\sqrt{2})}$ \\ \hline\hline
$k=7\pi /4$ & $a_{+}=14.6926...,\quad a_{-}=0.0256601...$ & $1-\frac{3}{%
\sqrt{2}}+\sqrt{\frac{3}{2}(2-2\sqrt{2})}$ \\ \hline\hline
\end{tabular}
\end{center}


\begin{thebibliography}{99}
\bibitem{KQ}  {\small Korff C 2003 \emph{J. Phys. A: Math. Gen}. \textbf{36}
5229-5266}

\bibitem{Lb67a}  {\small Lieb E H 1967 \emph{Phys. Rev.} \textbf{162}
162--172}

\bibitem{Lb67b}  {\small Lieb E H 1967 \emph{Phys. Rev. Lett.} \textbf{18}
1046--1048}

\bibitem{Lb67c}  {\small Lieb E H 1967 \emph{Phys. Rev. Lett.} \textbf{19}
108--110}

\bibitem{St67}  {\small Sutherland B 1967 \emph{Phys. Rev. Lett.} \textbf{19}
103--104}

\bibitem{DFM}  {\small Deguchi T, Fabricius K and McCoy B M 2001 \emph{J.
Stat. Phys.} \textbf{102} 701-736}

\bibitem{KM01}  {\small Korff C and McCoy B M 2001 \emph{Nucl. Phys}. B 
\textbf{618} [FS] 551-569}

\bibitem{Bx71}  {\small Baxter R J 1971 \emph{Phys. Rev. Lett}.\ \textbf{26}
193-228}

\bibitem{Bx72}  {\small Baxter R J 1972 \emph{Ann. Phys., NY} \textbf{70}
193--228}

\bibitem{Bx73a}  {\small Baxter R J 1973 \emph{Ann. Phys., NY} \textbf{76}
1--24}

\bibitem{Bx73b}  {\small Baxter R J 1973 \emph{Ann. Phys., NY} \textbf{76}
25--47}

\bibitem{Bx73c}  {\small Baxter R J 1973 \emph{Ann. Phys., NY} \textbf{76}
48--71}

\bibitem{BxBook}  {\small Baxter R J 1982 \emph{Exactly Solved Models in
Statistical Mechanics} (London: Academic Press)}

\bibitem{B31}  {\small Bethe H A 1931\ \emph{Z. Physik} \textbf{71} 205-226}

\bibitem{QISM}  {\small Fadeev L D, Sklyanin E K, Takhtajan L A 1979 Theor.
Math. Phys. \textbf{40} 194-220}

\bibitem{BLZ97}  {\small Bazhanov V, Lukyanov S and Zamolodchikov A 1997 
\emph{Comm. Math. Phys.} \textbf{190} 247--278}

\bibitem{BLZ99}  {\small Bazhanov V, Lukyanov S and Zamolodchikov A 1999 
\emph{Comm. Math. Phys.} \textbf{200} 297--324}

\bibitem{AF97}  {\small Antonov A and Feigin B 1997 \emph{Phys. Lett. B} 
\textbf{392} 115--122}

\bibitem{RW02}  {\small Rossi M and~Weston R~2002 \emph{J. Phys. A: Math.
Gen.} \textbf{35} 10015-10032}

\bibitem{RA89}  {\small Roche P and Arnaudon D 1989 \emph{Lett. Math. Phys}. 
\textbf{17} 295-300}

\bibitem{CK}  {\small De Concini C and Kac V 1990~\emph{Operator algebras,
unitary representations, enveloping algebras, and invariant theory\
(Progress in Mathematical Physics 92)\ }ed Connes A et al (Boston:
Birkh\"{a}user) pp 471}

\bibitem{CKP}  {\small De Concini~C, Kac~V and Procesi C 1992 \emph{J. Amer.
Math. Soc.} \textbf{5} 151-189}

\bibitem{A94}  {\small Arnaudon D 1994 \emph{Comm. Math. Phys}. \textbf{159}
175-194}

\bibitem{FM01a}  {\small Fabricius K and McCoy B M 2001 \emph{J. Stat. Phys.}
\textbf{103} 647-678}

\bibitem{BS90}  {\small Bazhanov V V and~Stroganov Yu G~1990 \emph{J. Stat.
Phys.\ }\textbf{59} 799-817}

\bibitem{Bx02}  {\small Baxter R J 2002 \emph{J. Stat. Phys.} \textbf{108}
1-48}

\bibitem{BA}  {\small Braak D and Andrei N 2001 \emph{J. Stat. Phys.} 
\textbf{105} 677-709}

\bibitem{FM01b}  {\small Fabricius K and McCoy B M 2002 \emph{MathPhys
Odyssey 2001} (\emph{Progress in Mathematical Physics 23}) ed Kashiwara M
and Miwa T (Boston: Birkhauser) pp 119}

\bibitem{Drin}  {\small Drinfeld V G~1988 \emph{Sov. Math. Dokl.} \textbf{36}
212-216}

\bibitem{CP}  {\small Chari V and Pressley A 1997 \emph{Representation Theory%
} \textbf{1} 280-328}

\bibitem{D02}  {\small Deguchi T 2002} {\small \emph{XXZ Bethe states as the
highest weight vector of the} }$sl_{2}${\small \ \emph{loop algebra at roots
of unity} cond-mat/0212217}

\bibitem{FM8v}  {\small Fabricius K and McCoy B M 2003 \emph{J. Stat. Phys.} 
\textbf{111} 323-337}
\end{thebibliography}
\end{document}